\documentstyle[psfig]{l-aa}  % LaTeX A&A Standard Fonts

\begin{document}
 
 \thesaurus{
	12(12.04.3;11.04.1;11.11.1)
	% Cosmology, distance scale,
	% Galaxies, distances and redshifts,
        % Galaxies, kinematics and dynamics
	}
 \title{Calibration of the Tully-Fisher relation in the field}
 
 \author{
S.~Rauzy 
}	
 
\offprints{S. Rauzy (rauzy@cpt.univ-mrs.fr)}
 
\institute{
ANPE and Centre de Physique Th\'eorique -
C.N.R.S., Luminy Case 907, F-13288 Marseille
 Cedex 9, France.
}

 \date{Received   / Accepted }
 \maketitle
 
 \begin{abstract}
A new technique for calibrating slope and relative zero-point
of Tully-Fisher like 
relations by using a sample of distant field galaxies is proposed.
Based on a
null-correlation approach (NCA),
the technique is insensitive to the presence of selection
effects on apparent magnitude $m$ and on log line-width
distance indicator $p$.
This interesting property
is used for discarding nearby galaxies of the 
observed sample. It is shown that such a subsampling
allows in effect to attenuate biases on calibration
parameters created by
the presence of radial peculiar velocities.


\end{abstract}

\section{Introduction}
\label{Introduction}

During this last decade, increasing efforts have been made 
attempting 
to extract informations concerning the distribution
of mass in the universe
from the radial peculiar velocity field of
distant galaxies (see for example
Aaronson et {al.} \cite{Aar86},
Lynden-Bell et { al.} \cite{Lyn88},
Bertschinger et { al.} \cite{Ber90},
Rauzy et { al.} \cite{Rau92},
Newsam et { al.} \cite{New95},
Rauzy et { al.} \cite{Rau95}).
Quantitative results based on peculiar velocity studies, such as
constraints on the value of the density parameter $\Omega_0$, can
nowadays be found in the literature (see Dekel \cite{Dek94}
for a review).
Reliability of such results is however closely related
to the various working hypotheses assumed throughout 
the successive
steps of the analysis.
\vspace{0.1 cm}

First of these steps 
is to obtain redshift-independent estimates
of galaxies distance. It is performed by using Tully-Fisher (TF)
like relations (Tully-Fisher \cite{Tul77} for spirals and
Faber-Jackson \cite{Fab76} for ellipticals). These
observed statistical
relations 
linearly correlate the absolute magnitude $M$ (or
similar quantity) of a galaxy
with an observable parameter $p$ ($p=\log V_{\rm rot}$ for spirals
and $p=\log \sigma$ for ellipticals).
Assuming that the TF relation has been correctly
calibrated, an estimate of the absolute magnitude $M$
is obtained by measuring the observable $p$.
Measurement of
the apparent magnitude $m$ (or
similar quantity) 
provides then with an estimate of the distance of the galaxy,
and comparing it with the redshift $z$ finally furnishes 
the deviation from the mean Hubble flow (i.e. the radial peculiar
velocity). 

The preliminar calibration step of the TF relation is
of crucial importance for kinematical analyses since
errors on the calibration parameters interpret indeed as
fictitious large-scale and coherent peculiar velocity
fields. 
Unfortunatly, selection effects in observation,
such as upper bound in apparent magnitude, bias the
estimates of the TF calibration parameters. Many studies
devoted to correct on these biases have been already
proposed (see for example
Schechter \cite{Sch80},
Bottinelli et al. \cite{Bot86},
Lynden-Bell et al. \cite{Lyn88},
Fouqu\'e et al. \cite{Fou90},
Hendry\&Simmons \cite{Hen90},
Teerikorpi \cite{Tee90},
Bicknell \cite{Bic92},
Triay et al. \cite{Tri94},
Willick \cite{Wil94},
Hendry\&Simmons \cite{Hen94},
Sandage \cite{San94},
Willick et al. \cite{Wil95},
Willick et al. \cite{Wil96},
Rauzy\&Triay \cite{Rau96},
Ekholm \cite{Ekh96},
Triay et al. \cite{Tri96}). 

Motivations leading to introduce a new calibration
technique are twofold.
First, bias correction requires generally a full 
description of the calibration sample
(i.e. the specific shapes of the observational selection function on $m$ and $p$
and of the luminosity function
have to be assumed). Since avaible samples are
often constituted of data inherited from various observational
programs, modelization of such characteristics still remains
a difficult problem.
The philosophy is herein to reduce as far as possible
the number of dubious assumptions made on these 
composite samples when deriving calibration parameters.
Second, it is not clear how existing calibration procedures
are affected by the presence of radial peculiar
velocities. The aim is herein to quantify, and if
possible to minimize, influences of peculiar
velocity fields on the estimates of the calibration
parameters.

In section \ref{dvestimate} is summarized the basic
statistical model describing Tully-Fisher like
relations. 
The new calibration technique
is presented
section \ref{calibrationNCA}.
Cumbersome calculations and proofs can be found
appendices A, B and C.
Potentialities of the method are illustrated appendix D where
NCA calibration of the Mathewson spirals field galaxies
sample is performed. 

\section{Distance and velocity estimates}
\label{dvestimate}
The Tully-Fisher like relations are based on an observed
linear correlation between the absolute magnitude $M$
(or similar quantity such as $-5\log_{10} D$ with $D$ the
linear diameter)
and the log line-width distance indicator $p$ of galaxies
($p \approx \log V_{\rm rot}$ for spirals and $p \approx
\log \sigma$ for ellipticals). They allow to estimate 
distance and
radial peculiar velocity of an individual galaxy 
from its measured apparent magnitude $m$ 
(or similar quantity such as $-5\log_{10} d$ with $d$ the
apparent diameter),
parameter $p$
and redshift $z$.
In this section, we recall in mind the basic statistical model
describing these relations
(see Triay et { al.} \cite{Tri94} or Rauzy\&Triay \cite{Rau96}
for details).

Regardless of the distance of the galaxy, selection effects in
observation and measurement errors, the theoretical probability
density ({\it pd}) in the $M$-$p$ plane reads :
\begin{equation}
\label{dpth}
dP_{1} = F(M,p)~dM\,dp
\end{equation}
The Direct (i.e. Forward) Tully-Fisher (DTF) relation
assumes that it exists a random variable
$\zeta^{D}$,
\begin{equation}
\label{zetad}
\zeta^{D} = \widetilde{M}(p) - M =
a^{D} p + b^{D} - M 
\end{equation}
statistically independent on $p$ such as the 
theoretical {\it pd} of Eq.
(\ref{dpth}) rewrites :
\begin{equation}
\label{dpthd}
dP_{1} = f_p(p)\,dp~
 g(\zeta^{D};0,\sigma_{\zeta}^{D})\,d\zeta^{D}  
\end{equation}
where 
$f_p(p)$
is the distribution function of the variable $p$ 
in the $M$-$p$ plane.
The random variable
$\zeta^{D}$ of zero mean and dispersion $\sigma_\zeta^{D}$
accounts for
the intrinsic scatter about the DTF straight line $\Delta_{\rm DTF}$
of zero-point $b^{D}$ and slope $a^{D}$. 
The distance modulus $\mu$ of an object reads :
\begin{equation}
\label{distancemodulus}
\mu= m - M
= 5 \log_{10} r +25 
\end{equation}
where $m$ is the apparent magnitude of the galaxy and
$r$ its distance in Mpc.
Regardless of measurement errors, the observed probability
density takes the following form :
\begin{equation}
\label{dpobs}
dP_{2}={{1
}\over{A_{2}}}
\phi(m,p)
~f_p(p)\,dp~g(\zeta^{D};0,\sigma_{\zeta}^{D})\,d\zeta^{D}\,
h(\mu)\,d\mu
\end{equation}
where
$\phi(m,p)$
is a selection function accounting for selection effects
in observation on $m$ and $p$,
$h(\mu)$
is the spatial distribution function of sources (along the line-of-sight)
and  $A_{2}=\int
\phi(m,p)
f_p(p)g(\zeta^{D};0,\sigma_\zeta^D)h(\mu)\,dp\,d\mu\,d\zeta^{D}$
is the normalisation factor warranting $\int dP_2 = 1$.
Under the three following assumptions :
\begin{itemize}
\item ${\cal H}$0) No measurement errors on $m$ and $p$ are
present. Particularly, corrections on galactic extinction
and on inclination effects are supposed valid.  
\item ${\cal H}$1) 
The function $g(\zeta^{D};0,\sigma_{\zeta}^{D})$ is gaussian.
\item ${\cal H}2$) Galaxies are homogeneously distributed in space,
which implies
that, whatever the line-of-sight direction, the distance
modulus distribution function reads
$h(\mu)=\exp [3\alpha\mu]$ with 
$\alpha={{\ln 10}\over{5}}$.
\end{itemize}
a statistical estimator
${\tilde r}$
generally adopted for the distance of a galaxy 
with measured $m$ and $p$ reads as
follows (see appendix \ref{Distanceestimate} for details,
or Lynden-Bell et al. \cite{Lyn88}
Landy\&Szalay \cite{Lan92}, Triay
et  al. \cite{Tri94}) :
\begin{equation}
\label{distanceestimator}
{\tilde r}=\exp[\alpha(m-a^{D}p-b^{D}-25)]\,\,\exp[{{7}\over{2}}\alpha^2
\sigma_\zeta^{D\,2}]
\end{equation}
where the term $\exp[{{7}\over{2}}\alpha^2 \sigma_\zeta^{D\,2}]$
accounts for a volume correction.
This unbiased\footnote{
The estimator ${\tilde r}$ given Eq. (\ref{distanceestimator})
is unbiased in the following sense.
Suppose a sample of $N$ galaxies 
homogeneously distributed in space and
with
the same measured $m$ and $p$. 
For $N$ large enough, the distances average
on the sample $\langle r \rangle$ will coincide with the
estimator ${\tilde r}$ (i.e. $\langle r \rangle \rightarrow
E(r \mid m,p)={\tilde r}$ where $E(r \mid m,p)$ is the
mathematical expectancy of $r$, given $m$ and $p$).
The statistics of
Eq. (\ref{distanceestimator}) is generally used for inferring 
the distance of individual galaxies. It amounts
to apply the above statistical formalism on a sample
containing only one object, which cannot be done without
ambiguousness.}   
distance estimator
does not depend on the selection function
$\phi(m,p)$ in $m$ and $p$ 
and on the specific shape of the distribution function
$f_p(p)$. 
Its accuracy $\Delta {\tilde r}$ is proportional to the distance
 estimate of the galaxy (see appendix \ref{Distanceestimate}) :
\begin{equation}
\label{raccuracy}
\Delta {\tilde r}={\tilde r}\,\sqrt{\exp[\alpha^2
\sigma_\zeta^{D\,2}]-1}
\end{equation}
The radial peculiar velocity $v$ of a galaxy, expressed in
km s$^{-1}$ with respect to the velocity frame in which the redshift
$z$ is measured, reads :
\begin{equation}
\label{velocity}
v=z-H_0\,r
\end{equation}
where $H_0$ is the Hubble's constant and $z$ is expressed
in km s$^{-1}$
units. 
Assuming that the above hypotheses hold,
the estimator
${\tilde v}$ of the radial peculiar velocity of a galaxy
with measured $m$, $p$ and $z$ reads thus as follows : 
\begin{equation}
\label{velocityestimator}
{\tilde v}=z-H_0\,{\tilde r}=
z-B\,\exp[\alpha(m-a^Dp)]\,\exp[{{7}\over{2}}\alpha^2
\sigma_\zeta^{D\,2}]
\end{equation}
where $H_0$ and 
$b^D$
have been merged into a single parameter
$B=H_0\exp[\alpha(-b^D-25)]$.
The accuracy 
$\Delta {\tilde v}$
of  this radial peculiar velocity estimator
${\tilde v}$ is :
\begin{equation}
\label{vaccuracy}
\Delta {\tilde v}=
H_0\, \Delta {\tilde r}=
H_0\,{\tilde r}\,\sqrt{\exp[\alpha^2
\sigma_\zeta^{D\,2}]-1}
\end{equation}

\section{Calibration using null-correlation approach}
\label{calibrationNCA}

Presence of the radial peculiar
velocities $v$ are included in the statistical modelization
by rewriting the density probability $dP_2$ of Eq.
(\ref{dpobs}) 
as follows :
\begin{equation}
\label{dpvelocity}
dP_3={{1
}\over{A_3}}
\phi(m,p)
f_p(p)dp\,g(\zeta^{D})d\zeta^{D}
h(\mu)d\mu
~f_v(v;{\bf x})dv
\end{equation}
where  
$f_v(v;{\bf x})$ is the distribution function of radial
peculiar velocities depending in general
on the spatial position
${\bf x}=(r \cos l \cos b, r \sin l \cos b, r \sin b)$ with 
$(l,b)$ the direction of the line-of-sight in
galactic coordinates.

The aim is herein to estimate the calibration parameters
of the DTF relation entering into 
the radial peculiar velocity estimator
${\tilde v}$ of Eq (\ref{velocityestimator}) (i.e. the DTF
slope
$a^D$ and DTF "zero-point" defined as $B^*= 
B\,\exp[{{7}\over{2}}\alpha^2
\sigma_\zeta^{D\,2}]$).
The calibration sample
is constituted of field galaxies selected along the same line-of-sight
of direction $(l,b)$ for which apparent magnitude $m$,
log line-width distance indicator $p$ and redshift $z$
are measured. It is assumed hereafter that the sample
is described by the density probability of Eq.  
(\ref{dpvelocity}) and
satisfies the following hypotheses :
\begin{itemize}
\item ${\cal H}$0) After appropriate corrections (galactic extinction,
inclination effects, ...), residual measurement errors can be neglected.
\item ${\cal H}$1) 
$g(\zeta^{D})$ is gaussian, i.e.
$g(\zeta^{D})=
g_G(\zeta^{D};0,\sigma_{\zeta}^{D})$.
\item ${\cal H}2$) 
The distance modulus distribution function reads
$h(\mu)=\exp [3\alpha\mu]$ with 
$\alpha={{\ln 10}\over{5}}$.
\item $\cal H$3)
The distribution of radial peculiar velocities $v$
is a gaussian of
mean $u$ and dispersion $\sigma_v$,
i.e.  $f_v(v;{\bf x})=g_{\rm G}(v;u,\sigma_v)$.
\end{itemize}
Since the calibration sample is constituted of galaxies lying
in the field, ${\cal H}2$ appears as a reasonable assumption.
Anyway, if the line-of-sight direction of the sample goes across a
physical cluster, nothing prevents us to discard
galaxies known to belong to the cluster.

Assumption ${\cal H}3$ implies that the radial peculiar velocity field
is not correlated with the distance $r$.
${\cal H}3$ is thus ruled out if flows such
like the Great Attractor are present along the
line-of-sight. On the other hand, ${\cal H}3$
is less restrictive than the pure Hubble's flow hypothesis
(i.e. $v=0$ everywhere). First, galaxies
may have a Maxwellian agitation
of velocity dispersion $\sigma_v$. 
Second, 
the mean radial peculiar velocity along the line-of-sight $u$
is not forced to zero. It allows to mimic the
following situation. Suppose that a whole-sky sample
is calibrated in a velocity frame of reference where
sampled galaxies are not globally at rest (say
that the sample has a bulk flow ${\bf u}=(u_x,u_y,u_z)$
in cartesian galactic coordinates with respect
to the velocity frame of reference).
The radial peculiar velocity of galaxies belonging
to the same line-of-sight of direction $(l,b)$ will
be shifted by 
${u}=u_x \cos l \cos b+u_y \sin l \cos b+u_z \sin b$.
Assumption ${\cal H}3$ in fact tolerates this
kind of situation.

\subsection{NCA calibration of the DTF slope ${\it a^D}$ }
\label{slopecalibration}  

Calibration of the DTF slope $a^D$ using null-correlation
approach (NCA) is based on the following remark.
For a calibration sample satisfying assumptions
${\cal H}0$,
${\cal H}1$,
${\cal H}2$ and
${\cal H}3$ with $u=\sigma_v=0$
(i.e. pure Hubble's flow hypothesis),
the variable $X=X(a^D)$ defined as : 
\begin{equation}
\label{XaD1}
X=X(a^D)=\alpha (m-a^D p) -\ln z
\end{equation}
is not correlated with $p$ (see appendix C
for proof) :
\begin{equation}
\label{CovpX1}
{\rm If}~~~ u=\sigma_v=0 ~~~
{\rm then}~~~
{\rm Cov}(p,X(a^D))= 0
\end{equation}
where ${\rm Cov}(p,X)=E(p\,X)-E(p)\,E(X)$ is the covariance
of variables $p$ and $X$. 
On the other hand,
a wrong value of the
DTF slope (i.e. $a_F=a^D+\Delta a$) correlates $p$ and
the random variable $X(a_F)=
\alpha (m-a_F\, p) -\ln z$ :
\begin{equation}
\label{correlationpXF1}
{\rm Cov}(p,X(a_F))~=~
\Delta C_0~=~
-\alpha \,\Delta a\, {\rm Cov}(p,p)
\ne~ 0 
\end{equation}
where 
${\rm Cov}(p,p)=E([p-E(p)]^2)=\Sigma(p)\,^2$ is the square of
the standard deviation 
of $p$. The null-correlation approach consists in to
adopt the value of the parameter $a$ verifying 
${\rm Cov}(p,X(a))=0$ as the correct estimate
of the DTF slope $a^D$ :
\begin{figure*}[htbp]
%\begin{figure*}
%\picplace{11.5 cm}
\centerline{\psfig{file=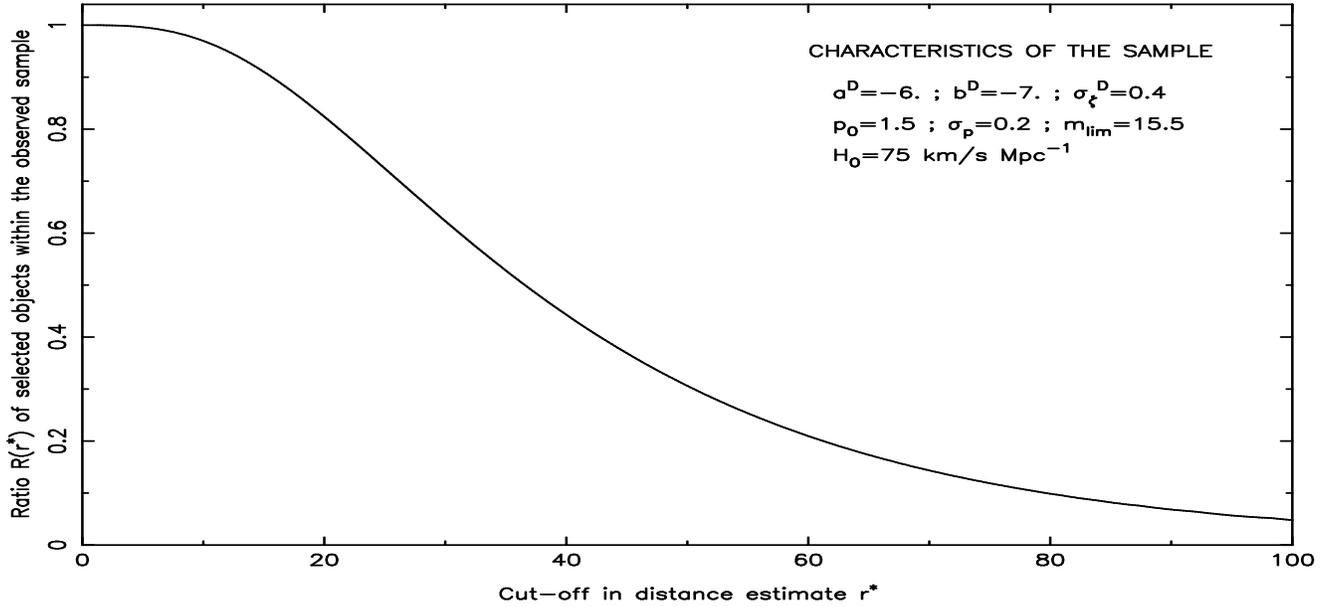,height=9. cm,width=\linewidth,angle=270}}
\caption[]{
Variation of the ratio $R(r^*)$ of selected object within the observed
sample with respect to the extra cut-off in distance estimate $r^*$.
}
\end{figure*}
%\begin{figure*}
\begin{figure*}[htbp]
%\picplace{10.5 cm}
\centerline{\psfig{file=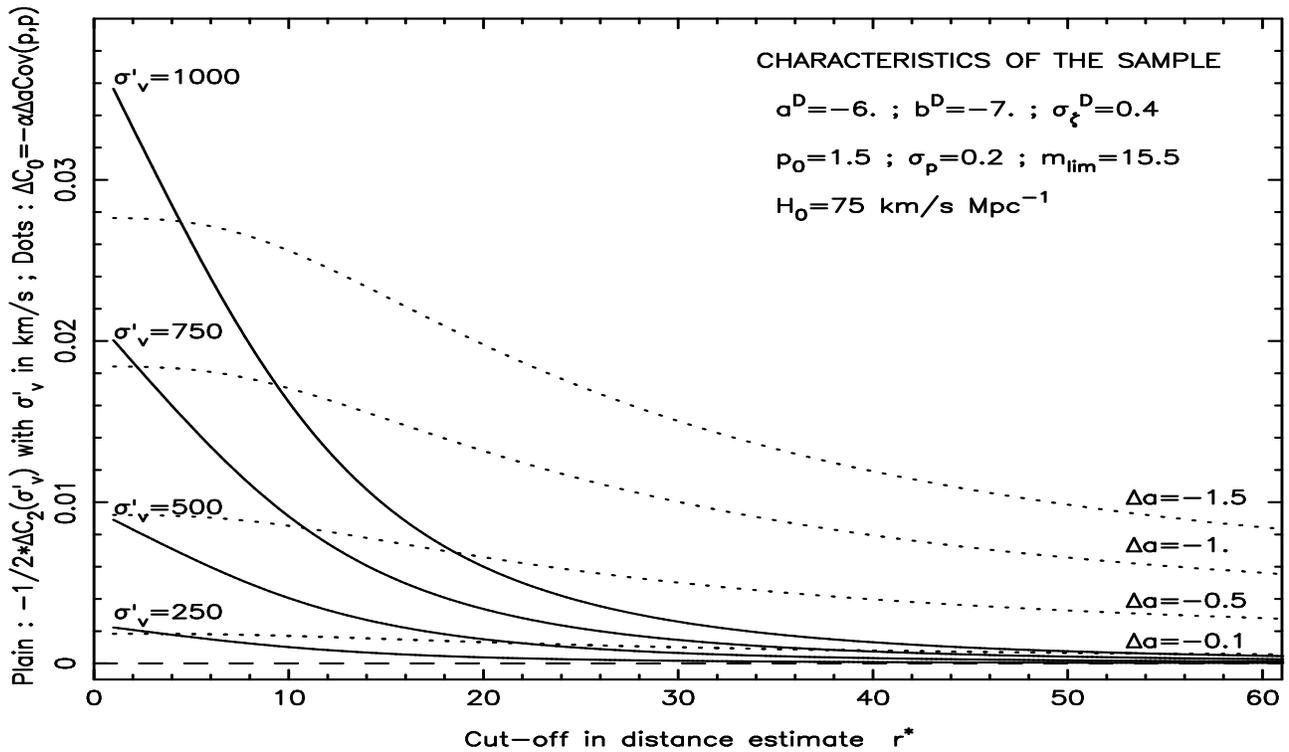,height=11. cm,width=\linewidth,angle=270}}
\caption[]{
Variation of the bias $\Delta C_2$ inferred by the presence
of a Maxwellian velocity agitation of dispersion $\sigma_v$
with respect to the extra cut-off in distance estimate $r^*$
($\sigma'_v=(\sigma_v^2+u^2)^{1/2}$).
}
\end{figure*}
\begin{equation}
\label{NCAslopeestimate}
{\rm NCA~ estimate~of~}a^D~:~~a~~
{\rm such}~~{\rm Cov}(p,X(a))~=~0
\end{equation}
which gives in practice :
\begin{equation}
\label{NCAslopeestimate1}
a^{\rm NCA}~ = ~
\frac{{\rm Cov}(p,m)}
{{\rm Cov}(p,p)}
~-~
\frac{1}
{\alpha}
\frac{{\rm Cov}(\,p\,,\,\ln z\,)}
{{\rm Cov}(p,p)}
\end{equation}
Note that in the
case of pure Hubble's flow, the NCA is a quite
robust technique for calibrating the slope of the
DTF relation. It furnishes indeed an unbiased estimate
of $a^D$ whatever the selection effects $\phi(m,p)$ 
on $m$ and $p$ which affect the observed sample,
and whatever the specific shape of the theoretical
distribution function $f_p(p)$ of the variable $p$
(see appendix C). 

Unfortunatly, 
the presence of radial peculiar velocities such as
the ones assumed in ${\cal H}3$ with $u\ne 0$ or
$\sigma_v \ne 0$ biases the NCA estimate of the slope
$a^D$ since Eq. (\ref{CovpX1}) rewrites in this case 
(see appendix C) :
\begin{equation}
\label{correlationpX2}
{\rm Cov}(p,X(a^D))~=~
-~{\rm Cov}(p,\ln[1+\frac{v}{H_0\,r}])
~\ne~ 0 
\end{equation}
However, the magnitude of this bias
can be attenuated by
selecting only distant galaxies of the observed sample. It is not
surprising since the term 
$\ln[1+v/(H_0r)]$ in Eq. (\ref{correlationpX2})
becomes negligible for distances $r$ large enough.
Such a subsampling can be performed by
discarding galaxies of the observed sample which have
a distance estimate 
${\tilde r}$
smaller than
a given $r^*$. It corresponds to introduce an extra selection
function $\psi(m,p)$ defined as follows :
\begin{equation}
\label{extraselection1}
\psi(m,p)=1~~{\rm if}~~{\tilde r}
\, \ge\, r^*~~~;~~~\psi(m,p)=0~~{\rm otherwise.}
\end{equation}
where ${\tilde r}$ is given Eq. (\ref{distanceestimator}).
Introducing  this extra selection effect does not
alterate the result  obtained Eq. (\ref {CovpX1}) since
this property is insensitive to the specific shape adopted
for the
selection function on $m$ and $p$.

In order to evaluate amplitude of the bias appearing
Eq. (\ref{correlationpX2}) and its variation with
respect to the cut-off in distance estimate $r^*$,
calculations have been performed on a synthetic
sample characterized as follows :
\begin{itemize}
\item 
${\cal H}0$, ${\cal H}1$, ${\cal H}2$
and ${\cal H}3$ are satisfied by the sample.
\item 
DTF slope $a^D=-6.$
\item 
DTF zero-point $b^D=-7.$
\item 
DTF intrinsic scatter $\sigma_\zeta^D=0.4$
\item 
Hubble's constant $H_0=75$ km s$^{-1}$ Mpc$^{-1}$
\item 
The distribution function of $p$ is a gaussian of mean
$p_0=1.5$ and dispersion $\sigma_p=0.2$
\item 
The selection effects 
in observation restrict to a cut-off in apparent magnitude
$m_{\rm lim}=15.5$
\end{itemize}
The two dominant terms of the bias created by the presence
of radial peculiar
velocities  
have been calculated
(see appendix C for details) : 
\begin{equation}
\label{Covexpansion1}
{\rm Cov}(p,\ln[1+\frac{v}{H_0\,r}]) =
\Delta C_1 -\frac{1}{2} \Delta C_2 +\circ (\Delta C_2)
\end{equation}
where the analytical expressions of
$\Delta C_1$ and $\Delta C_2$ in function of $u$,
$\sigma_v$ and $r^*$ can be found
Eq. (\ref{deltac12}).
The amplitude of these terms has to be compared with $\Delta C_0
=-\alpha \Delta a \,{\rm Cov}(p,p)$ appearing in Eq. (\ref{correlationpXF1})
when a wrong value of the DTF slope is adopted.
Expression of $\Delta C_0$ in function of $\Delta a$ and $r^*$
is given Eq. (\ref{deltac0}).

Figure 1 shows variation of $R(r^*)$, i.e.
the ratio of selected objects within the observed sample,
with respect to the cut-off in distance estimate
$r^*$ (analytical expression of $R(r^*)$ is
given Eq. (\ref{ratiomuetoile})). On one hand,
high values of $r^*$ are required in order to minimize
as far as possible amplitude of the bias created by
radial peculiar velocities. On the other hand,
accuracy $\sigma_a$ of the NCA slope estimate
depends on the size of the selected subsample
(i.e. $\sigma_a \propto R(r^*)^{-1/2}$ due to
the intrinsic statistical fluctuations affecting
the sample). Since theses two features are
indeed competitive (i.e. $R(r^*)$ decreases when $r^*$ increases,
compromise on the optimal value of $r^*$ has to be
chosen with regard to the specific
characteristics of the data sample under consideration.

The variations of term $\Delta C_2$, i.e. the contribution
of a Maxwellian agitation of velocity dispersion
$\sigma_v$ to the bias of Eq. 
(\ref{Covexpansion1}), are illustrated Fig. 2.
Influence of $\Delta C_2$ on the NCA estimate
of the DTF slope $a^D$ given Eq. 
(\ref{NCAslopeestimate}) is obtained by comparing
$\Delta C_2$ with the contribution 
of $\Delta C_0
=-\alpha \Delta a \,{\rm Cov}(p,p)$ to the covariance
of Eq. (\ref{correlationpXF1}).
For example, if the mean radial peculiar velocity
along the line-of-sight $u$ is zero,
the magnitude of the bias on the NCA estimate of $a^D$
created by a velocity field of dispersion
$\sigma_v = 1000$ km s$^{-1}$ is greater than
$\Delta a = -1.5$ for $r^*=0$, 
falls to 
$\Delta a = -1.$ for $r^* \approx 10$ (i.e. $R(r^*) \approx 0.95$), 
equalizes
$\Delta a = -0.5$ for $r^* \approx 20$ (i.e. $R(r^*) \approx 0.8$) 
and finally falls below
$\Delta a = -0.1$ for $r^* \approx 60$ (i.e. $R(r^*) \approx 0.2$).
Figure 2 reveals two important features.
\\- 
If nearby galaxies are not discarded from the observed sample,
the presence of a Maxwellian velocity agitation for
galaxies contaminates strongly the NCA estimate of the DTF
slope $a^D$ (a velocity dispersion of 
$\sigma_v = 500$ km s$^{-1}$
induces a bias on $a^D$ of magnitude
$\Delta a \approx -0.5$). 
\\- 
This bias can be reendered arbitrarily small by selecting
only distant galaxies by means of the extra cut-off in
distance estimate $r^*$.

\begin{figure*}
%\begin{figure*}[htbp]
%\picplace{22. cm}
\centering \psfig{file=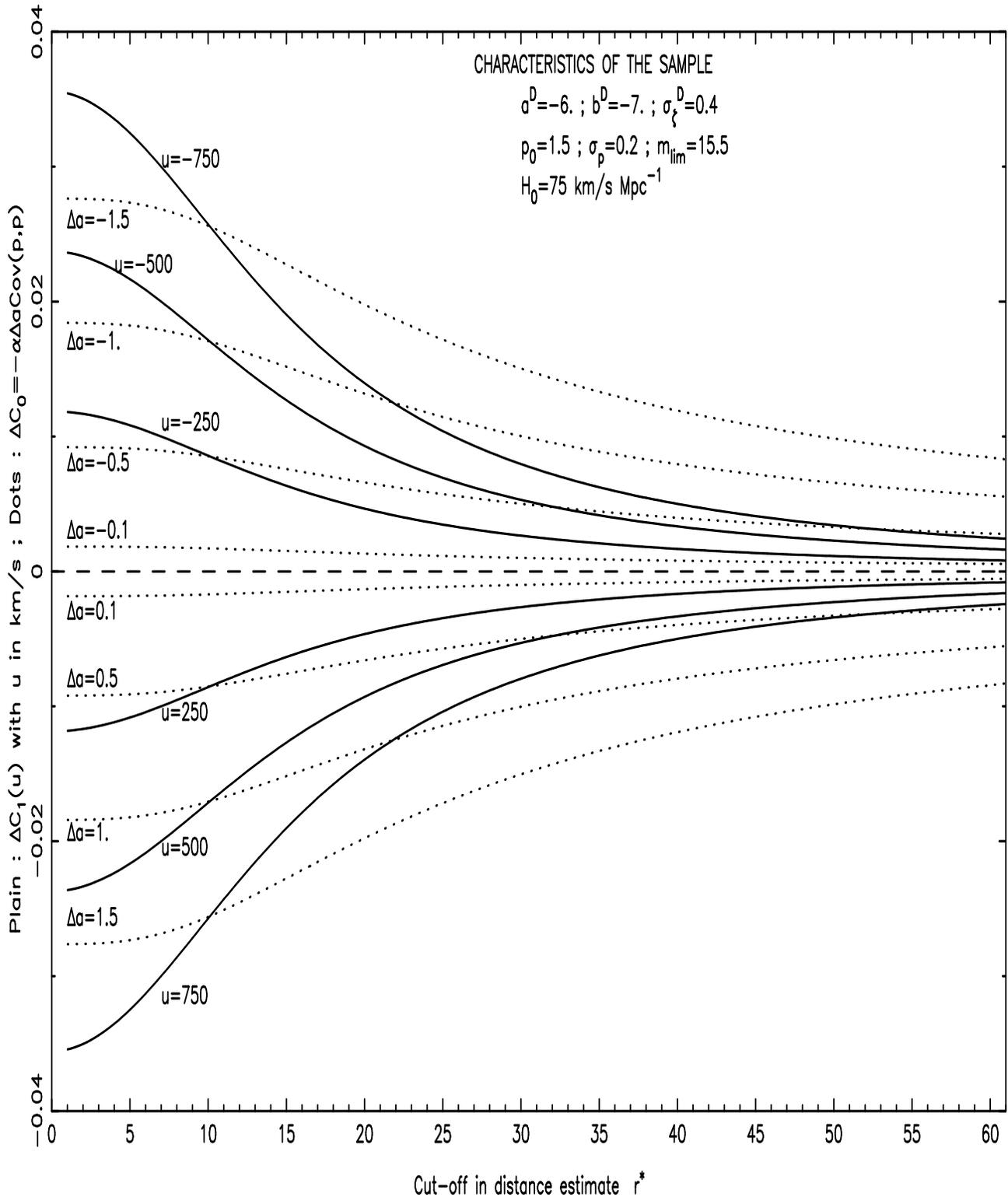,height=22. cm,width=17.5 cm,angle=270}
\caption[]{
Variation of the bias $\Delta C_1$ inferred by the presence
of a constant velocity $u$ along the line-of-sight
with respect to the extra cut-off in distance estimate $r^*$.
}
\end{figure*}

Figure 3 shows the variations of $\Delta C_1$, i.e.
the dominant term of the bias entering Eq.
(\ref{Covexpansion1}) due to the presence of a mean 
radial peculiar velocity along the line-of-sight $u$,
with respect to $r^*$.
A careful analysis of Fig. 3 leads to the three
following remarks.

The term $\Delta C_1$ decreases in function of $r^*$
less rapidly than $\Delta C_2$.
For comparable values of $\Delta C_1$ and $\Delta C_2$
at $r^*=0$, say for
$u = -750$ km s$^{-1}$
and $\sigma'_v = 1000$ km s$^{-1}$,
$\Delta C_1$ bias falls to 
$\Delta a \approx -1.5$ at $r^* =10 $ (to
be compared with 
$\Delta a \approx -1.$ for $\Delta C_2$ bias), 
to $\Delta a \approx -1.$ at $r^* =20 $ 
($\Delta a \approx -0.5$ for $\Delta C_2$) and finally to 
$\Delta a \approx -0.5$ at $r^* =60 $ 
($\Delta a \approx -0.1$ for $\Delta C_2$).
It thus turns out that a particular attention 
has to be paid in priority to the presence of
constant velocity fields.

As the existence of constant velocities strongly biases
the NCA estimate of $a^D$ (at $r^*=0$, 
$\Delta a \approx -1.3$ for 
$u = -500$ km s$^{-1}$), discarding nearby 
galaxies by means of distance estimate selection appears
as a quite crucial step. Note however that the situation
is not so stringent for samples affected by bulk flow.
Since the $\Delta C_1$ bias is antisymmetric with respect
to $u$ (see Fig. 3), the bias on the NCA estimate of $a^D$
cancels in average if the opposite line-of-sight
direction is also considered. By scanning the sky by line-of-sight
directions, this interesting symmetry allows indeed to
detect bulk flows already at the level of the calibration step.

Finally, some upper bounds on the $a^D$ bias created by
large-scale coherent velocity fields can be extracted from
analysis of Fig. 3. Suppose that the line-of-sight
of the calibration sample points toward the direction
of a Great Attractor, located at a distance estimate of
${\tilde r}=40$ Mpc and creating back-side infall velocities,
say of amplitude 
$u = -500$ km s$^{-1}$ at
${\tilde r}=50$ Mpc and slowly decreasing at larger distances.
The bias on the NCA estimate of $a^D$ will be necessarily
smaller than the $\Delta C_1$ bias for 
$u = -500$ km s$^{-1}$
(i.e. at $r^*=50$, $| \Delta a | < 0.5$). This property
is closely related to the subsampling in distance estimate
allowed by the null-correlation approach. The peculiar
velocity field, whatever its specific form, becomes
negligible compared to the mean Hubble's flow as long
as the cut-off in distance estimate $r^*$ is large enough.
In this case the null-correlation approach furnishes unbiased
estimate of the DTF slope $a^D$.

\subsection{NCA calibration of the DTF "zero-point" ${\it B^*}$ }
\label{zeropointcalibration}  

Assuming that the DTF slope $a^D$ has been correctly calibrated
by means of the technique presented section
\ref{slopecalibration} or others calibration procedures,
the null-correlation approach is herein used for
calibrating the remaining calibration parameters entering
into the radial peculiar velocity estimator ${\tilde v}$
of Eq. (\ref{velocityestimator}). 
For this purpose, 
this equation is
rewritten as follows :
\begin{equation}
\label{Betoile}
{\tilde v}=
z - H_0 \,{\tilde r} =
z-B^*\,\exp[\alpha(m-a^Dp)]
\end{equation}
where $B^*=H_0\exp[\alpha(-b^D-25)]
\,\exp[{{7}\over{2}}\alpha^2
\sigma_\zeta^{D\,2}]$ depends on 
$H_0$,  
$b^D$ and $\sigma_\zeta^D$.
For a calibration sample satisfying assumptions
${\cal H}0$,
${\cal H}1$,
${\cal H}2$ and
${\cal H}3$, 
the radial peculiar velocity estimator
${\tilde v}=
{\tilde v}(a^D,B^*)$
is not correlated with $p$ (see appendix B
for proof) :
\begin{equation}
\label{Covpv1}
{\rm Cov}(p,{\tilde v})~=~
{\rm Cov}(p,{\tilde v}(a^D,B^*))~=~ 0
\end{equation}
On the other hand,
a wrong value of the $B^*$ parameter 
(i.e. $B_F=B^*+\Delta B^*$)
correlates $p$ and the random variable
${\tilde v}(a^D,B_F)=
z-B_F\,\exp[\alpha(m-a^Dp)]$ :
\begin{equation}
\label{correlationpvF1}
{\rm Cov}(p,
{\tilde v}(a^D,B_F))~=~
-\,\frac{\Delta B^*}{B^*}\,
{\rm Cov}(p,
H_0\,{\tilde r})~ \ne ~0
\end{equation}
which does not vanish since 
selection effects on apparent magnitude $m$ correlate
variables $p$ and ${\tilde r}$ (i.e. for 
selected galaxies, observable $p$ increases in average
with the distance estimate ${\tilde r}$).
Assuming that the DTF slope $a^D$ has been correctly
calibrated, the NCA estimate of the "zero-point" $B^*$ 
is then defined as :
\begin{equation}
\label{NCAzeropointestimate}
{\rm NCA~ estimate~of~}B^*:~\beta~~
{\rm such}~~{\rm Cov}(p,{\tilde v}(a^D, \beta))=0
\end{equation}
which gives in practice :
\begin{equation}
\label{NCAzeropointestimate1}
B_*^{\rm NCA}~ = ~
\frac{{\rm Cov}\left(\,p\,,\,\exp [\alpha(m-a^Dp)]\,\right)}
{{\rm Cov}(p,z)}
\end{equation}
NCA estimate of $B^*$ is clearly robust. It is
insensitive to observational selection effects
on $m$ and $p$, specific shape of the
luminosity function $f_p(p)$, constant velocity
field and Maxwellian agitation of galaxies (see appendix B).
Note however that presence of
non-constant large scale velocity fields (such GA flow
for example) biases NCA estimate of $B^*$. Unfortunatly
the subsampling procedure in distance estimate previously
proposed is not efficient for dealing with
this kind of biases since the contribution of
the peculiar velocity $v$ to ${\rm Cov}(p,z)$ entering Eq.
(\ref{NCAzeropointestimate1}) does not decrease with the distance $r$.

If one want to express the distance estimator ${\tilde r}$
of Eq. (\ref{distanceestimator}) in true distance units (Mpc),
the Hubble's constant $H_0$ has to be estimated (i.e.
$H_0 \,{\tilde r} = B^* \, \exp [\alpha (m -a^D p)]$). 
For this purpose, estimates of the parameters $B^*$ 
---by using NCA calibration for example---, 
DTF zero-point $b^D$
---using primary distance indicators---
and DTF intrinsic dispersion $\sigma_\zeta^D$
---in galaxies clusters for example---
are required (i.e. $H_0= 
B^*\exp[\alpha(b^D+25)]
\,\exp[-{{7}\over{2}}\alpha^2
\sigma_\zeta^{D\,2}]$).
If $B^*$ is estimated with an error of $\delta B^*$,
$b^D$ and $\sigma_\zeta^D$ with errors of
$\delta b^D$ 
and $\delta \sigma_\zeta^D$ 
respectively,
the dominant term of the relative error on the $H_0$ estimate
reads :
\begin{equation}
\label{H0accuracy}
\frac{\delta H_0}{H_0}
~\approx
~ \frac{\delta B^*}{B^*}~+
~\alpha\,\delta b^D~ -
~{{7}}\alpha^2
\,\sigma_\zeta^D \,\delta \sigma_\zeta^{D}
\end{equation}

\section{Conclusion}
\label{Conclusion}

A new technique for calibrating Tully-Fisher like relations
was proposed. 
Based on a null-correlation approach,
this calibration procedure is efficient when galaxies
constituting the calibration sample are homogeneously distributed in space.
The NCA technique is thus adequate for calibrating
a sample of field galaxies for example.

In a first step was introduced
the random variable $X(a)=\alpha(m-a\, p) - \ln z$
dependent on a slope parameter $a$ and on the
observed apparent
magnitude $m$, log line-width distance indicator $p$
and redshift $z$. 
In the case of pure Hubble's flow (i.e. radial peculiar velocities
are null everywhere), it was shown that variables $p$ and
$X(a)$ are not correlated if and only if parameter $a$
equals the slope $a^D$ of the Direct TF relation.
The NCA estimate of the DTF slope $a^D$ was defined
as the value of $a$ such that correlation between $p$
and $X(a)$ vanishes.
This estimator of $a^D$ was found particularly robust since it
does not depend on the selection effects $\phi(m,p)$ on
$m$ and $p$ which affect the observed calibration sample
and on the specific shape of the luminosity
function.

Influences of radial peculiar velocities was investigated
in a second step. It was shown that the presence
of a peculiar velocity field biases the NCA estimate of $a^D$.
A procedure which 
consists in discarding nearby galaxies
of the observed sample was introduced. 
Such subsampling is achieved by adding an
extra selection function $\psi(m,p)$ in distance estimate ${\tilde r}$. The
magnitude of the bias and its variations in function of
the cut-off in distance estimate $r^*$ have been analysed
on a synthetic sample and on the Mathewson spirals field galaxies
sample (see appendix D). The proposed
subsampling procedure
looks fairly efficient for minimizing bias on the
NCA estimate of the DTF slope $a^D$ created by the presence
of radial peculiar velocities.

In the third step, calibration of the "zero-point" 
parameter $B^*$ entering the definition of the velocity
estimator ${\tilde v}$ was
investigated. It was shown that
the variables $p$ and 
${\tilde v}(a^D,\beta)=z-\beta \,\exp[\alpha(m-a^Dp)]$
are not correlated if and only if parameter $\beta$
equals the "zero-point" $B^*$.
The NCA estimate of the second calibration parameter 
$B^*$ was thus defined
as the value of $\beta$ such that correlation between $p$
and ${\tilde v}(a^D,\beta)$ vanishes. 
This $B^*$ estimator is robust. It is
insensitive to observational selection effects
on $m$ and $p$ and to the specific shape of the
luminosity function. 
Moreover, NCA estimate of $B^*$ is not biased by the presence
of a Maxwellian velocity agitation nor by the 
existence of bulk flow.

\begin{acknowledgements}
This work is one of the achievements of a long range program
focusing on the statistical modelization of TF like relations,
launched five years
ago by Roland Triay. He is kindly thanked for constructive discussions.
The hospitality
of the Centre de Physique Th\'eorique of Luminy
is recognized. 
St\'ephane Rauzy is cheerfully thanked for a generous financial support.
\end{acknowledgements}

\onecolumn
\appendix
\section{The distance estimator 
${\rm {\tilde {\it r}}}$}
\label{Distanceestimate}  

The probability density $dP_{m_o,p_o}$ describing
a sample of galaxies
with the same observed $m_o$ and $p_o$ can be derived from
Eq. (\ref{dpobs}) by using conditional probability :
\begin{equation}
\label{dpmopo}
dP_{m_o,p_o}={{1
}\over{A_{m_o,p_o}}}~
\delta(m-m_o)\,\delta(p-p_o)~
\phi(m,p)
~f_p(p)\,dp~g(\zeta^{D};0,\sigma_{\zeta}^{D})\,d\zeta^{D}\,
h(\mu)\,d\mu
\end{equation}
where $\delta$ is the Dirac distribution 
(i.e. $\int \delta(x-x_0)\,f(x)\,dx = f(x_0)$) and 
the normalisation factor $A_{m_o,p_o}$ reads :
\begin{equation}
\label{Amopo}
A_{m_o,p_o}=
\int \delta(m-m_o)\,\delta(p-p_o)~
\phi(m,p)
~f_p(p)\,dp~g(\zeta^{D};0,\sigma_{\zeta}^{D})\,d\zeta^{D}\,
h(\mu)\,d\mu
\end{equation}
If the two following hypotheses are verified by the sample :
\begin{itemize}
\item $\cal H$1)
$g(\zeta^{D};0,\sigma_{\zeta}^{D})$ is gaussian,
i.e. $g \equiv g_{\rm G}$ with
$g_{\rm G}(x;x_0,\sigma) =
(\sqrt{2\pi}\,\sigma)^{-1}\exp[-(x-x_0)^2/(2\sigma^2)]$ .
\item $\cal H$2)
Galaxies are homogeneously distributed
along the line-of-sight, i.e.
$h(\mu)=\exp [3\alpha\mu]$ with 
$\alpha={{\ln 10}\over{5}}$ . 
\end{itemize}
the successive integrations over $m$, $p$ and $\zeta^D$ give for
the normalisation factor $A_{m_o,p_o}$ :
\begin{equation}
\label{Amoporeduced}
A_{m_o,p_o}=
\phi(m_o,p_o)
~f_p(p_o)~\exp[3 \alpha \,(m_o-a^D p_o -b^D)]~\exp[{{9}\over{2}} \alpha^2
\sigma_\zeta^D\,^{2}]
\end{equation}
where the properties (\ref{property}) and
(\ref{gaussian}a)
of gaussian functions have been used :
\begin{equation}
\label{property}
\exp [\lambda\,x]~ g_{\rm G}(x;x_0,\sigma) =
\exp [\lambda\,(x_0+\lambda\,{\sigma^2}/{2})]~ g_{\rm G}(x;x_0+
\lambda \, \sigma^2,\sigma)
\end{equation}
\begin{equation}
\label{gaussian}
\int g_{\rm G}(x;x_0,\sigma)~dx =1~~~~{\rm (a)}~~;~~~~~
\int x~g_{\rm G}(x;x_0,\sigma)~dx =x_0~~~~{\rm (b)}~~;~~~~~
\int x^2~g_{\rm G}(x;x_0,\sigma)~dx =\sigma^2+x_0^2~~~~{\rm (c)}
\end{equation}
By using property (\ref{property}) and Eq. (\ref{Amoporeduced}),
the distribution of distance modulus $\mu$ for the sample rewrites
thus :
\begin{equation}
\label{dPmu}
dP_{m_o,p_o}=
~g_G(\mu;
m_o-a^D p_o -b^D + 3 \alpha \,
\sigma_{\zeta}^{D}\,^{2},
\sigma_{\zeta}^{D})
\,d\mu
\end{equation}
Properties (\ref{property})
and (\ref{gaussian}a)
imply that the mathematical expectancy
$E(r)$ of the distance $r=\exp [\alpha\,(\mu -25)]$
reads :
\begin{equation}
\label{distanceexpectancy}
E(r)=
\int r~dP_{m_o,p_o} =
\exp[\alpha(m_o-a^{D}p_o-b^{D}-25)]\,\,\exp[{{7}\over{2}}\alpha^2
\sigma_\zeta^{D\,2}]
\end{equation}
This quantity is generally chosen as 
distance estimator ${\tilde r}$
of individual galaxy with measured $m_o$ and $p_o$. 
Note however that others estimators could be used,
such as the most likely value of the distance $r$ which does not coincide
with $E(r)$ since the probability density function ({\it pdf})
of the
variable $r$ is lognormal, and so
not symmetric around its mean. 
The accuracy 
$\Delta {\tilde r}$ of the distance estimator ${\tilde r}$
can be derived from Eqs. (\ref{dPmu},\ref{distanceexpectancy})
by using properties (\ref{property}) and (\ref{gaussian}a) :
\begin{equation}
\label{rtildeaccuracy}
\Delta {\tilde r}=
\sqrt{ E((r-{\tilde r})^2)}=
{\tilde r}~\sqrt{\exp[\alpha^2
\sigma_\zeta^{D\,2}]-1}
\end{equation}

\section{Correlation between ${\rm {\it p}}$ and 
${\rm {\tilde {\it v}}}$}
\label{Correlationpv}  

Hereafter, the correlation between $p$ and
the velocity estimator ${\tilde v}$ given Eq. (\ref{Betoile})
is calculated 
assuming the following hypothesis : 
\begin{itemize}
\item $\cal H$3)
Along a line-of-sight, the distribution of radial peculiar velocities $v$
does not depend on distances $r$ and is a gaussian of 
mean $u$ and dispersion $\sigma_v$,
i.e.  $f_v(v;{\bf x})=g_{\rm G}(v;u,\sigma_v)$ .
\end{itemize}
Under assumptions 
$\cal H$1, $\cal H$2 and $\cal H$3, the probability density
$dP_3$ of Eq. (\ref{dpvelocity}) rewrites :
\begin{equation}
\label{dp3}
dP_3={{1
}\over{A_3}}
\phi(m,p)
~f_p(p)\,dp~g_{\rm G}(\zeta^{D};0,\sigma_\zeta^D)\,d\zeta^{D}\,
\exp[3\alpha\mu]\,d\mu
~g_{\rm G}(v;u,\sigma_v)\,dv
\end{equation}
In order to calculate each term involved in the covariance
of $p$ and ${\tilde v}$, i.e. ${\rm Cov}(p,{\tilde v})=
E(p\, {\tilde v})-E(p)\,E({\tilde v})$, the velocity estimator
${\tilde v}$ given Eq. (\ref{velocityestimator}) is rewritten
in terms of variables $v$, $\zeta^D$ and $\mu$ :
\begin{equation}
\label{vtildevmu}
{\tilde v}=v+
H_0\,\exp[\alpha(\mu-25)]\,\,\left 
[1-\exp[-\alpha \zeta^D]\,\exp[{{7}\over{2}}\alpha^2
\sigma_\zeta^{D\,2}]\right ]
\end{equation}
Since $u$ is a constant, 
${\rm Cov}(p,{\tilde v}-u)=
E(p \,({\tilde v}-u))-E(p)\,E({\tilde v}-u)=
{\rm Cov}(p,{\tilde v})$.
Integrating over the variable $v$ gives for
the three quantities $E(p)=\int p~ dP_3$,
$E({\tilde v}-u)=
\int ({\tilde v}-u)~dP_3$ and 
$E(p\,({\tilde v}-u))=
\int p\,({\tilde v}-u)~dP_3$ :
\begin{equation}
\label{B1}
E(p) =
{{1
}\over{A_3}}
\int \phi(m,p)
~p\,f_p(p)dp\,g_{\rm G}(\zeta^{D};0,\sigma_\zeta^D)d\zeta^{D}\,
h(\mu)d\mu
\end{equation}
\begin{equation}
\label{B2}
E({\tilde v}-u)=
{{1
}\over{A_3}}
\int \phi(m,p)
~f_p(p)dp\,g_{\rm G}(\zeta^{D};0,\sigma_\zeta^D)d\zeta^{D}\,
h(\mu)d\mu
\,H_0\exp[\alpha(\mu-25)]\left 
[1-\exp[-\alpha \zeta^D]\exp[{{7}\over{2}}\alpha^2
\sigma_\zeta^{D\,2}]\right ]
\end{equation}
\begin{equation}
\label{B3}
E(p\,({\tilde v}-u))=
{{1
}\over{A_3}}
\int \phi(m,p)
~p \,f_p(p)dp\,g_{\rm G}(\zeta^{D};0,\sigma_\zeta^D)d\zeta^{D}\,
h(\mu)d\mu
\,H_0\exp[\alpha(\mu-25)]\left 
[1-\exp[-\alpha \zeta^D]\exp[{{7}\over{2}}\alpha^2
\sigma_\zeta^{D\,2}]\right ]
\end{equation}
where properties
(\ref{gaussian}a) and
(\ref{gaussian}b) have been used.
Replacing $h(\mu)$ by $\exp[3\alpha \mu]
=\exp[3 \alpha(m-a^{D}p-b^{D}+\zeta^D)]$ and using definition
of $B$ given Eq. (\ref{velocityestimator}),
Eqs. (\ref{B1},\ref{B2},\ref{B3})
expressed in terms of $p$, $m$ and $\zeta^D$ read :
\begin{equation}
\label{B4}
E(p) =
{{1
}\over{A_3}}
\int \phi(m,p)
~p\,f_p(p)
\exp[3 \alpha(m-a^{D}p-b^{D})]
\,dm\,dp 
~\times
\int
\exp[3\alpha \zeta^D]
\,g_{\rm G}(\zeta^{D};0,\sigma_\zeta^D)\,d\zeta^{D}
\end{equation}
\begin{equation}
\label{B5}
E({\tilde v}-u)=
{{B
}\over{A_3}}
\int \phi(m,p)
\,f_p(p)
\exp[4 \alpha(m-a^{D}p-b^{D})]
\,dm\,dp 
~\times
C
\end{equation}
\begin{equation}
\label{B6}
E(p\,({\tilde v}-u))=
{{B
}\over{A_3}}
\int \phi(m,p)
~p\,f_p(p)
\exp[4 \alpha(m-a^{D}p-b^{D})]
\,dm\,dp 
~\times
C
\end{equation}
where the constant $C$ is defined as follows :
\begin{equation}
\label{B7}
C=
\int 
\left 
[\exp[4\alpha \zeta^D]-\exp[3\alpha \zeta^D]\,\exp[{{7}\over{2}}\alpha^2
\sigma_\zeta^{D\,2}]\right ]
\,g_{\rm G}(\zeta^{D};0,\sigma_\zeta^D)\,d\zeta^{D}
\end{equation}
By using properties (\ref{property})
and (\ref{gaussian}a), it is found that integral of Eq.
(\ref{B7}) vanishes. It thus implies that $C=0$, 
$E({\tilde v}-u)=0$,
$E(p\,({\tilde v}-u))=0$,
${\rm Cov}(p,{\tilde v}-u)=0$ and finally
${\rm Cov}(p,{\tilde v})=0$.
This result does not depend on the shape
of the selection function $\phi(m,p)$ in $m$ and $p$,
on the theoretical distribution function $f_p(p)$ of
the variable $p$ and on the mean radial peculiar
velocity $u$ along the line-of-sight.
For a sample of galaxies satisfying hypotheses 
$\cal H$1, $\cal H$2 and $\cal H$3, it thus turns
out that the observable $p$ is not correlated
with the velocity estimator
${\tilde v}$
as long as the model parameters entering into
${\tilde v}={\tilde v}(a^D,b^D,\sigma_\zeta^D,H_0)$
via Eq. (\ref{velocityestimator}) are correct\footnote{
Note however that the converse does not hold
since the Hubble's constant $H_0$, DTF zero-point $b^D$
and intrinsic DTF dispersion $\sigma_\zeta^D$ degenerate
into a single parameter $B^*= H_0 \exp[-\alpha(b^D+25)]\,
\exp[{{7}\over{2}}\alpha^2
\sigma_\zeta^{D\,2}]$.}. On the other hand, if a wrong value
is assumed for one of these calibration parameters, 
say for example that
$a^D$, $b^D$ 
and $\sigma_\zeta^D$ are correct but not the Hubble's constant 
$H_F=H_0+\Delta H_0$,
the covariance between $p$ and 
$v_F={\tilde v}(a^D,b_D,\sigma_\zeta^D,H_F)$ gives :
\begin{equation}
\label{correlationpvF}
{\rm Cov}(p,v_F)~=~
{\rm Cov}(p,
{\tilde v}-\Delta H_0 {\tilde r})
~=~
0\,-\Delta H_0\,
{\rm Cov}(p,
{\tilde r})~ \ne ~0
\end{equation}
which does not vanish since 
existing selection effects on apparent magnitude $m$ correlate
variables $p$ and ${\tilde r}$ (i.e. for selected
galaxies, observable
$p$ increases in average with the distance estimator  
${\tilde r}$). This point motivates the use of a null-correlation
approach\footnote{
It can be proven similarly that ${\rm Cov}(m,{\tilde v})~=~0$ .}
in order to calibrate Tully-Fisher like relations.

\section{Correlation between ${\rm {\it p}}$ and 
${\rm {\it X}}$}
\label{CorrelationpX}  

In this appendix, the reliability of null-correlation approach  
in order to estimate the slope $a^D$ of the DTF relation
is discussed.
In a first step, 
it is shown that
the non-observable
variable $X_0=X_0(a^D)=\alpha (m-a^D p) -\ln [H_0\,r]$ is not
correlated with $p$
as long as hypotheses ${\cal H}1$ and ${\cal H}2$ are satisfied.
Particularly, this property does not depend on the selection function on
$m$ and $p$.
In a second step, the correlation between $p$ and the observable 
variable $X$ :
\begin{equation}
\label{XaD}
X=X(a^D)=\alpha (m-a^D p) -\ln z=X_0-\ln[1+\frac{v}{H_0\,r}]
\end{equation}
is investigated.
The presence of radial peculiar velocities such as
the ones assumed in ${\cal H}3$ indeed correlates 
$p$ and $X$.  However, it is shown that
the amplitude of this effect can be weakened by
selecting only distant galaxies of the sample. It is not
surprising since the term 
$\ln[1+v/(H_0r)]$ in Eq. (\ref{XaD})
becomes negligible for distances $r$ large enough.
Such a subsampling can be performed by
discarding galaxies of the observed sample which have
a distance estimate 
${\tilde r}$ less than
a given $r^*$. It corresponds to introduce an extra selection
function $\psi(m,p)$ defined as follows :
\begin{equation}
\label{extraselection}
\psi(m,p)=1~~{\rm if}~~{\tilde r}=\exp[\alpha(m-a^{D}p-b^{D}-25)]\,
\exp[{{7}\over{2}}\alpha^2
\sigma_\zeta^{D\,2}]\, \ge\, r^*~~~;~~~\psi(m,p)=0~~{\rm otherwise.}
\end{equation}
Note that introducing this extra selection effect does not
alterate the result
${\rm Cov}(p,X_0)=0\,$. On the other hand,
a wrong value of the
DTF slope (i.e. $a_F=a^D+\Delta a$) correlates $p$ and
$X(a_F)=
\alpha (m-a_F\, p) -\ln z$ :
\begin{equation}
\label{correlationpXF}
{\rm Cov}(p,X(a_F))~=~
{\rm Cov}(p,X_0(a_F))+
{\rm Cov}(p,-\ln[1+\frac{v}{H_0\,r}])~=~
-\alpha \Delta a~{\rm Cov}(p,p)-
{\rm Cov}(p,\ln[1+\frac{v}{H_0\,r}])~
\ne~ 0 
\end{equation}
where 
${\rm Cov}(p,p)=E([p-E(p)]^2)=\Sigma(p)\,^2$ is the square of
standard deviation 
of $p$. The null-correlation approach thus consists to
adopt the value of the parameter $a$ verifying 
${\rm Cov}(p,X(a))=0$ as the correct estimate
of the slope $a^D$ of the DTF relation.

\vspace{0.2 cm}
It is proven below that 
${\rm Cov}(p,X_0=\alpha (m-a^D p) -\ln [H_0\,r])=0$ as long
as assumptions ${\cal H}1$ and ${\cal H}2$ are satisfied
by the sample. Rewriting the non-observable variable $X_0$ in
terms of $m$, $p$ and $\mu$ gives $X_0= 
\alpha (m-a^D p -\mu +25) -\ln H_0$. In terms of $\zeta^D$, it
reads $X_0=-\alpha \zeta^D -\ln B$ where
$B=H_0\exp[\alpha(-b^D-25)]$ was previously defined
Eq. (\ref{velocityestimator}). Since $\alpha$ and $\ln B$ are
constants, 
${\rm Cov}(p,X_0)=
-\alpha \,{\rm Cov}(p,\zeta^D)$. The null-correlation between
$p$ and $\zeta^D$ can be shown by replacing
$h(\mu)$ by $\exp[3\alpha \mu]
=\exp[3 \alpha(m-a^{D}p-b^{D}+\zeta^D)]$ and using the probability 
density $dP_2$ of Eq. (\ref{dpobs})
expressed in terms of $p$, $m$ and $\zeta^D$ :
\begin{equation}
\label{C4}
E(p) =
{{1
}\over{A_2}}~Q_1 \times Q_2 =
{{1
}\over{A_2}}
\int \phi(m,p)
~p\,f_p(p)
\exp[3 \alpha(m-a^{D}p-b^{D})]
\,dm\,dp 
~\times
\int
\exp[3\alpha \zeta^D]
\,g_{\rm G}(\zeta^{D};0,\sigma_\zeta^D)\,d\zeta^{D}
\end{equation}
\begin{equation}
\label{C5}
E(\zeta^D) =
{{1
}\over{A_2}}~Q_3 \times Q_4 =
{{1
}\over{A_2}}
\int \phi(m,p)
\,f_p(p)
\exp[3 \alpha(m-a^{D}p-b^{D})]
\,dm\,dp 
~\times
\int
\zeta^D\,\exp[3\alpha \zeta^D]
\,g_{\rm G}(\zeta^{D};0,\sigma_\zeta^D)\,d\zeta^{D}
\end{equation}
\begin{equation}
\label{C6}
E(p\,\zeta^D) =
{{1
}\over{A_2}}~Q_1 \times Q_4 =
{{1
}\over{A_2}}
\int \phi(m,p)
~p\,f_p(p)
\exp[3 \alpha(m-a^{D}p-b^{D})]
\,dm\,dp 
~\times
\int
\zeta^D\,\exp[3\alpha \zeta^D]
\,g_{\rm G}(\zeta^{D};0,\sigma_\zeta^D)\,d\zeta^{D}
\end{equation}
where the normalisation factor $A_2$ reads :
\begin{equation}
\label{C7}
A_2 =
Q_3 \times Q_2 =
\int \phi(m,p)
\,f_p(p)
\exp[3 \alpha(m-a^{D}p-b^{D})]
\,dm\,dp 
~\times
\int
\exp[3\alpha \zeta^D]
\,g_{\rm G}(\zeta^{D};0,\sigma_\zeta^D)\,d\zeta^{D}
\end{equation}
It thus turns out that
${\rm Cov}(p,\zeta^D)=
E(p\,\zeta^D) -E(p)\,E(\zeta^D) =
(A_2)^{-2}\,[A_2 \, Q_1 \, Q_4 - Q_1 \, Q_2 \times Q_3\, Q_4]
=0$ , which implies 
${\rm Cov}(p,X_0)=0\,$.
This result\footnote{
It can be proven similarly that ${\rm Cov}(m,X_0)~=~0$ .}
is insensitive to the shape
of the selection function $\phi(m,p)$ in $m$ and $p$ and
to the theoretical distribution function $f_p(p)$ of
the variable $p$ . 
\vspace{0.2 cm}

The term 
${\rm Cov}(p,\ln[1+{v}/(H_0\,r)])$ entering into Eq.
({\ref{correlationpXF}) does not vanish since $p$ is correlated
with distance $r$. However,
its amplitude can be reendered arbitrarily small
by adding an extra selection effect such as the one mentioned
Eq. (\ref{extraselection}). This feature is illustrated
by the following example.
It is assumed that :
\begin{itemize}
\item 
Hypotheses ${\cal H}1$, ${\cal H}2$
and ${\cal H}3$ are satisfied by the sample.
\item 
The distribution function of $p$ is a gaussian of mean
$p_0$ and dispersion $\sigma_p$
(i.e. $f_p(p)=
g_{\rm G}(p;p_0,\sigma_p)\,$).
\item 
The selection effects 
in observation restrict to a cut-off in apparent magnitude
(i.e. $\phi(m,p) \equiv \phi_m(m)=\theta (m_{\rm lim}-m)$ where
$\theta$ is the Heaveside distribution function,
$\theta (x) = 1$ if $x \ge 0$ and $\theta (x) = 0$ otherwise). 
\end{itemize}
The subsampling is performed by using the extra selection
function $\psi(m,p)$ proposed Eq.
(\ref{extraselection}). By introducing $\mu^*$ such that
$r^*=\exp[\alpha(\mu^*-25)]\,
\exp[{{7}\over{2}}\alpha^2
\sigma_\zeta^{D\,2}]$, $\psi$ rewrites 
$\psi(m,p)=\theta(
[m-a^{D}p-b^{D}]-\mu^*)$. Adding this extra selection function
$\psi(m,p)$,
the probability density
$dP_3$ of Eq. (\ref{dpvelocity}) reads : 
\begin{equation}
\label{dp4}
dP_4={{1
}\over{A_4}}
\,\theta (m_{\rm lim}-m)\,
\theta(
[m-a^{D}p-b^{D}]-\mu^*)
\,g_{\rm G}(p;p_0,\sigma_p)
\,dp~g_{\rm G}(\zeta^{D};0,\sigma_\zeta^D)\,d\zeta^{D}\,
\exp[3\alpha \mu]\,d\mu
~g_{\rm G}(v;u,\sigma_v)\,dv
\end{equation}
Since the probability density $dP_4$ describing the sample is
fully specified, calculations of quantities of interest can
be performed analytically or by using numerical simulations.
Hereafter, cumbersome intermediate calculations have been
deliberately omitted and we just furnish the ultimate 
analytical expressions of these quantities.

In order to evaluate the amplitude of the term
${\rm Cov}(p,\ln[1+{v}/(H_0\,r)])$ entering Eq.
({\ref{correlationpXF}),
$\ln[1+{v}/(H_0\,r)]$ is expanded in Taylor's series up to
order 2 :
\begin{equation}
\label{lnexpansion}
\ln ( 1 + \frac{v}{H_0\, r} )~ =~
\frac{v}{H_0\, r}~ -
\frac{1}{2} 
\frac{v^2}{H_0^2\, r^2}~  +~
\circ \left ( \frac{v^2}{H_0^2\, r^2} \right )
\end{equation}
It implies that  the covariance may be expanded as follows :
\begin{equation}
\label{Covexpansion}
{\rm Cov}(p,\ln[1+\frac{v}{H_0\,r}]) =
\Delta C_1 -\frac{1}{2} \Delta C_2 +\circ (\Delta C_2) =
{\rm Cov}(p,\frac{v}{H_0\,r}) -
\frac{1}{2}{\rm Cov}(p,\frac{v^2}{H_0^2 r^2}) +
\circ \left ( {\rm Cov}(p,\frac{v^2}{H_0^2 r^2}) \right )
\end{equation}
The calculations give for the terms $\Delta C_1$ and $\Delta C_2$ :
\begin{equation}
\label{deltac12}
\Delta C_1 = \frac{u}{H_0}\,\left (\frac{J[\mu^*;2]}{I[\mu^*;3]}-
\frac{J[\mu^*;3]}{I[\mu^*;3]}
\frac{I[\mu^*;2]}{I[\mu^*;3]} \right )~~~~~~{\rm (a)}~~~~;~~~~
\Delta C_2 = \frac{u^2+\sigma_v^2}{H_0^2}
\,\left (\frac{J[\mu^*;1]}{I[\mu^*;3]}-
\frac{J[\mu^*;3]}{I[\mu^*;3]}
\frac{I[\mu^*;1]}{I[\mu^*;3]} \right )~~~~~~{\rm (b)}
\end{equation}
where the functions $I[\mu^*;N]$ and $J[\mu^*,N]$ are defined
Eq. (\ref{I}) and Eq. (\ref{J}). For a given cut-off in
distance estimate $\mu^*$, the contribution of $\Delta C_1$
and $\Delta C_2$ to ${\rm Cov}(p,X)$ 
has to be compared with $\Delta C_0
=-\alpha \Delta a {\rm Cov}(p,p)$ appearing in Eq. (\ref{correlationpXF})
when a wrong value of the DTF slope is adopted.
This term $\Delta C_0$ reads :
\begin{equation}
\label{deltac0}
\Delta C_0= -\alpha \, \Delta a\, {\rm Cov}(p,p) =
 -\alpha \, \Delta a\, 
\left (
\frac{H[\mu^*]}{I[\mu^*;3]}-
\frac{J[\mu^*;3]^2}{I[\mu^*;3]^2} \right )
\end{equation}
where the function $H[\mu^*]$ is defined Eq. (\ref{Ep2}) and
$I[\mu^*;3]$ and $J[\mu^*,3]$ 
Eqs. (\ref{I},\ref{J}). Finally, the last quantity of interest
is 
the ratio $R(\mu^*)$ of selected objects within the observed
sample. This ratio is equal to the normalisation
factor $A_4(\mu^*)=I(\mu^*;3)$ divided by
$A_4(\mu^* \rightarrow -\infty)$, i.e. the normalisation
factor of the sample when no extra selection effect is
present ($r^*=0$ or $\mu^* \rightarrow -\infty$).
It reads in function of the extra cut-off in distance estimate
$\mu^*$ introduced Eq. (\ref{dp4}) as follows :
\begin{equation}
\label{ratiomuetoile}
R(\mu^*)=
{\cal M}_0( \omega[\mu^*;3] )\,-\,
{\cal M}_0( \omega[\mu^*;0] )\,
\exp [3 \alpha (
\mu^*+ a^D p_0+b^D
-m_{\rm lim}+
\frac{3}{2} \alpha
a^D\,^2 \sigma_p^2) ]
\end{equation}
where 
$\omega[\mu^*;N]$ is defined Eq. (\ref{calmnmuetoile}) and
the 
${\cal M}_0(x)$ function is introduced Eq. (\ref{integralegaussian}).

\vspace{0.4 cm}
\noindent NOTATIONS :
\vspace{0.2 cm}

\noindent 
$I[\mu^*;N]$ and $J[\mu^*,N]$, function of the extra cut-off
$\mu^*$ and of an integer $N$, are defined as follows :
\begin{equation}
\label{I}
I[\mu^*;N]=K[N]\,\left (
{\cal M}_0(\omega[\mu^*;N])\,-\,
{\cal M}_0( \omega[\mu^*;0] )\,
\exp [N \alpha\, (
\mu^*+ a^D p_0+b^D
-m_{\rm lim}-
\frac{N}{2} \alpha
a^D\,^2 \sigma_p^2) ] \right )
\end{equation}
\begin{equation}
\label{J}
\begin{tabular}{ll}
$J[\mu^*;N]=$ & $ + K[N]\,(\,
\sigma_p\, {\cal M}_1(\omega[\mu^*;N])+
(p_0-N\,\alpha a^D \sigma_p^2)\,{\cal M}_0(\omega[\mu^*;N])\,)$ \\
\\
$\,$ & $ - K[N]\,
(\,\sigma_p\,{\cal M}_1( \omega[\mu^*;0] )+p_0\,
{\cal M}_0( \omega[\mu^*;0] )\,)\,
\exp [N \alpha (
\mu^*+ a^D p_0+b^D
-m_{\rm lim}-
\frac{N}{2} \alpha
a^D\,^2 \sigma_p^2) ] $
\end{tabular}
\end{equation}
where $K(N)$, function of an integer $N$, is :
\begin{equation}
\label{K}
K[N]=\frac{1}{N\,\alpha}\,\exp[N\,\alpha\,(
m_{\rm lim}-a^D p_0-b^D-25+\frac{N}{2}\,\alpha(
a^D\,^2 \sigma_p^2 + \sigma_\zeta^D\,^2))]
\end{equation}
${\cal M}_0(x)$,
${\cal M}_1(x)$ and
${\cal M}_2(x)$ functions reads :
\begin{equation}
\label{integralegaussian}
N(t)=
 g_{\rm G}(t;0,1)~~;~~~
{\cal M}_0(x)=\int_x^{+\infty} N(t)\,dt ~~;~~~
{\cal M}_1(x)=\int_x^{+\infty} t\, N(t)\,dt ~~;~~~
{\cal M}_2(x)=\int_x^{+\infty} t^2\,N(t)\,dt 
\end{equation}
and  $\omega[\mu^*;N]$, function of the extra cut-off $\mu^*$
and of an integer $N$, is :
\begin{equation}
\label{calmnmuetoile}
\omega[\mu^*;N]=
\frac{m_{\rm lim}-a^D p_0-b^D-\mu^*+N\,\alpha\,
a^D\,^2 \sigma_p^2}{a^D \sigma_p} 
\end{equation}
The function $H[\mu^*]$, involved in the calculation of 
${\rm Cov}(p,p)$, is defined as follows :
$$
\frac{H[\mu^*]}{K[3]}= 
-(\sigma_p^2{\cal M}_2( \omega[\mu^*;0] )+2\sigma_p p_0
{\cal M}_1( \omega[\mu^*;0] )+
p_0^2
{\cal M}_0( \omega[\mu^*;0] )\,)\,
\exp [3 \alpha (
\mu^*+ a^D p_0+b^D
-m_{\rm lim}+
\frac{3}{2} \alpha
a^D\,^2 \sigma_p^2) ] $$
\begin{equation}
\label{Ep2}
~~~~~~~~~~~~~~\,\,+
\sigma_p^2 {\cal M}_2(\omega[\mu^*;3])+
2\sigma_p(p_0-3\alpha a^D \sigma_p^2){\cal M}_1(\omega[\mu^*;3])+
(p_0-3\alpha a^D \sigma_p^2)^2{\cal M}_0(\omega[\mu^*;3]) 
\end{equation}

\begin{figure*}
%\begin{figure*}[htbp]
%\picplace{22. cm}
{\hbox{
\psfig{file=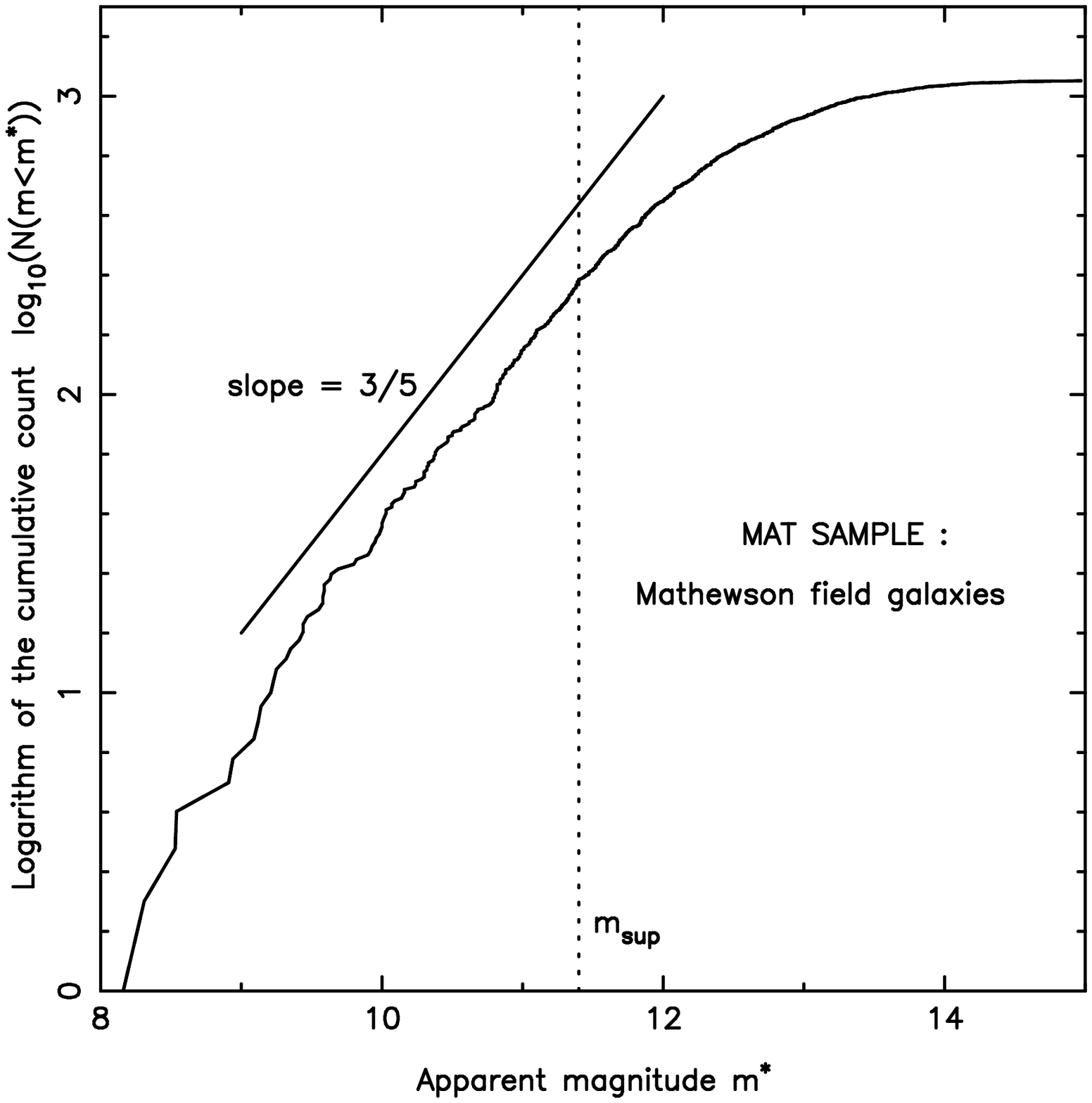,height=7.5 cm,width=8.7 cm,angle=270}
\psfig{file=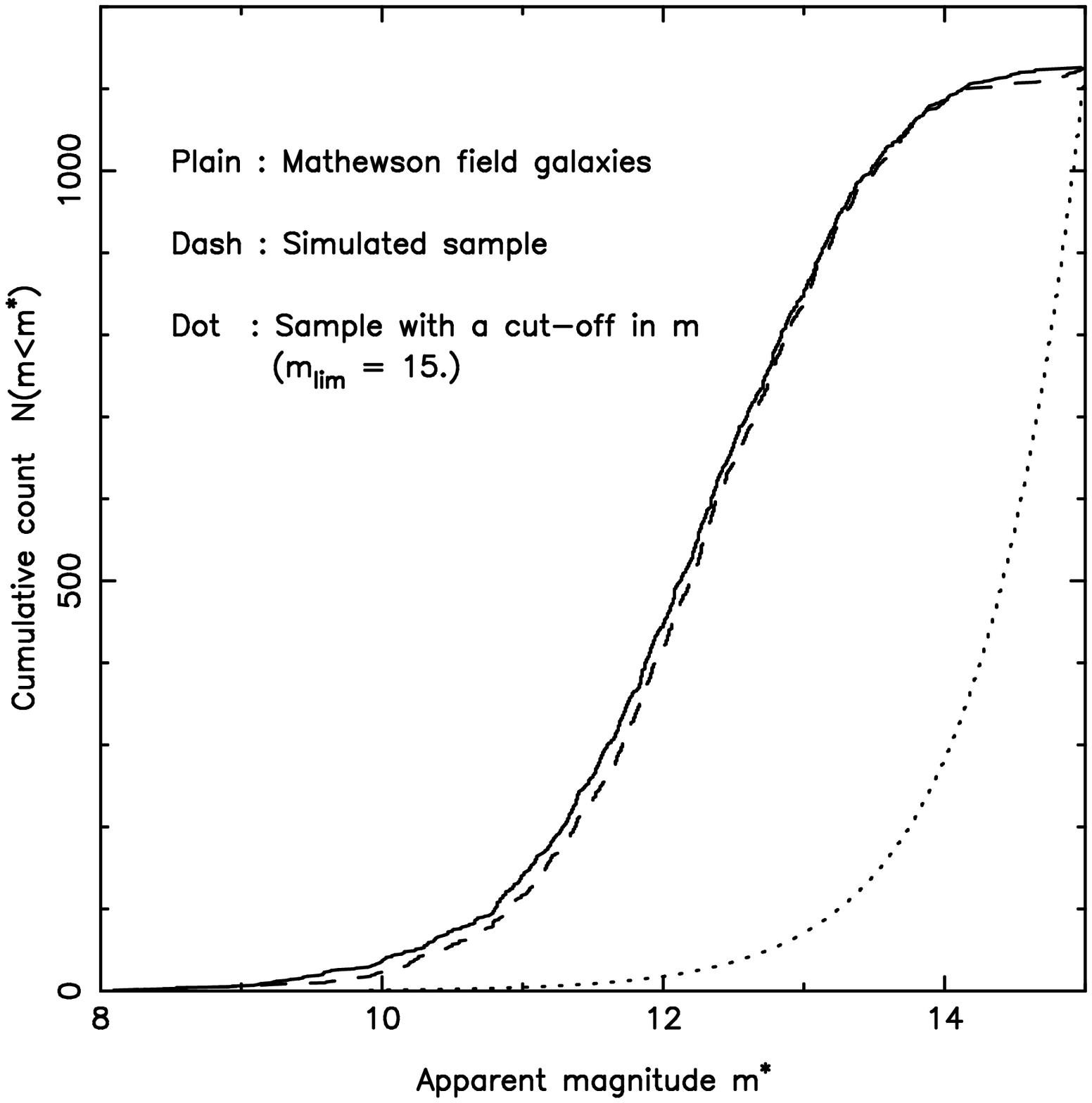,height=7.5 cm,width=8.7 cm,angle=270}
}}
{\hbox{
\psfig{file=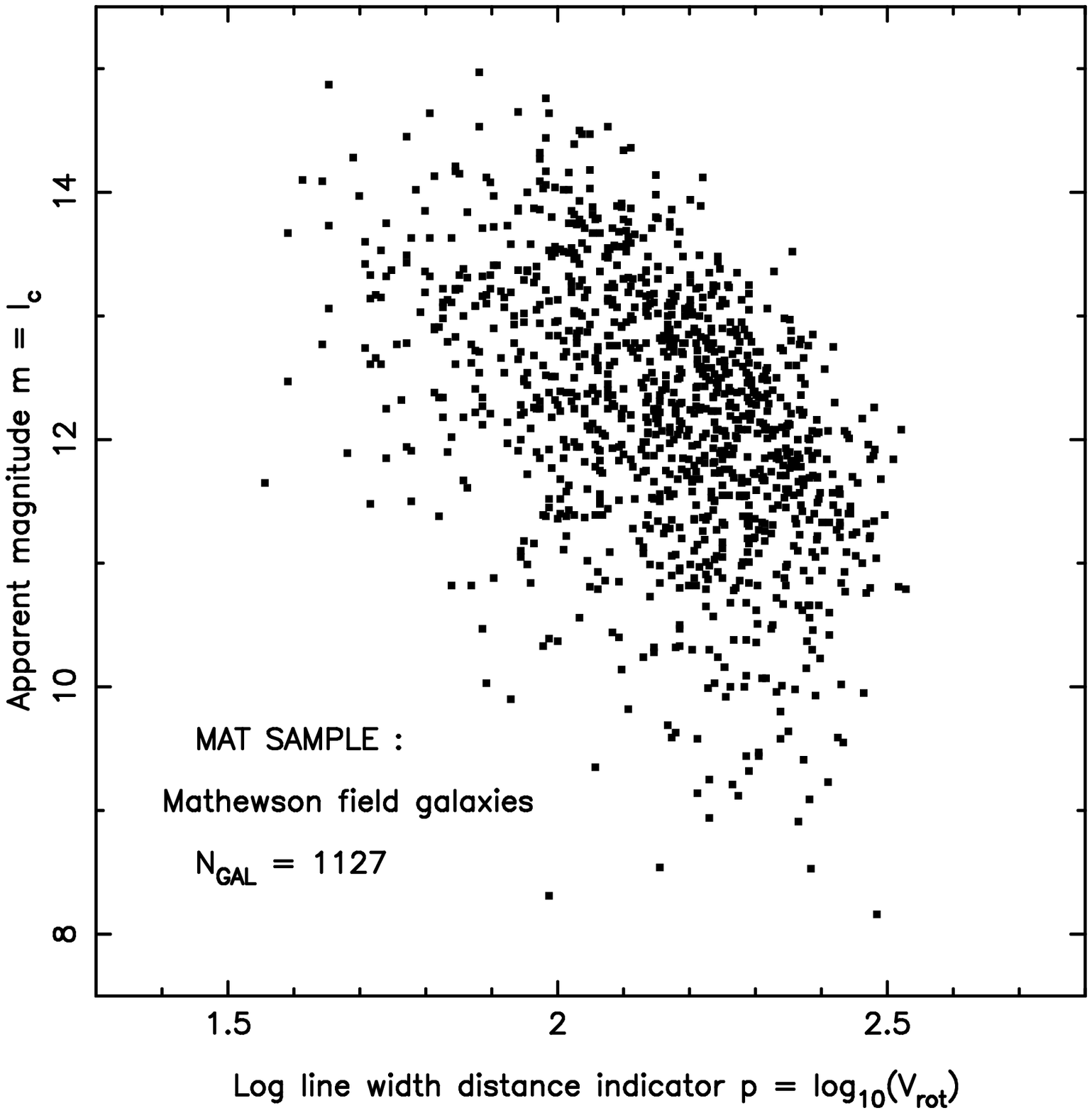,height=7.5 cm,width=8.7 cm,angle=270}
\psfig{file=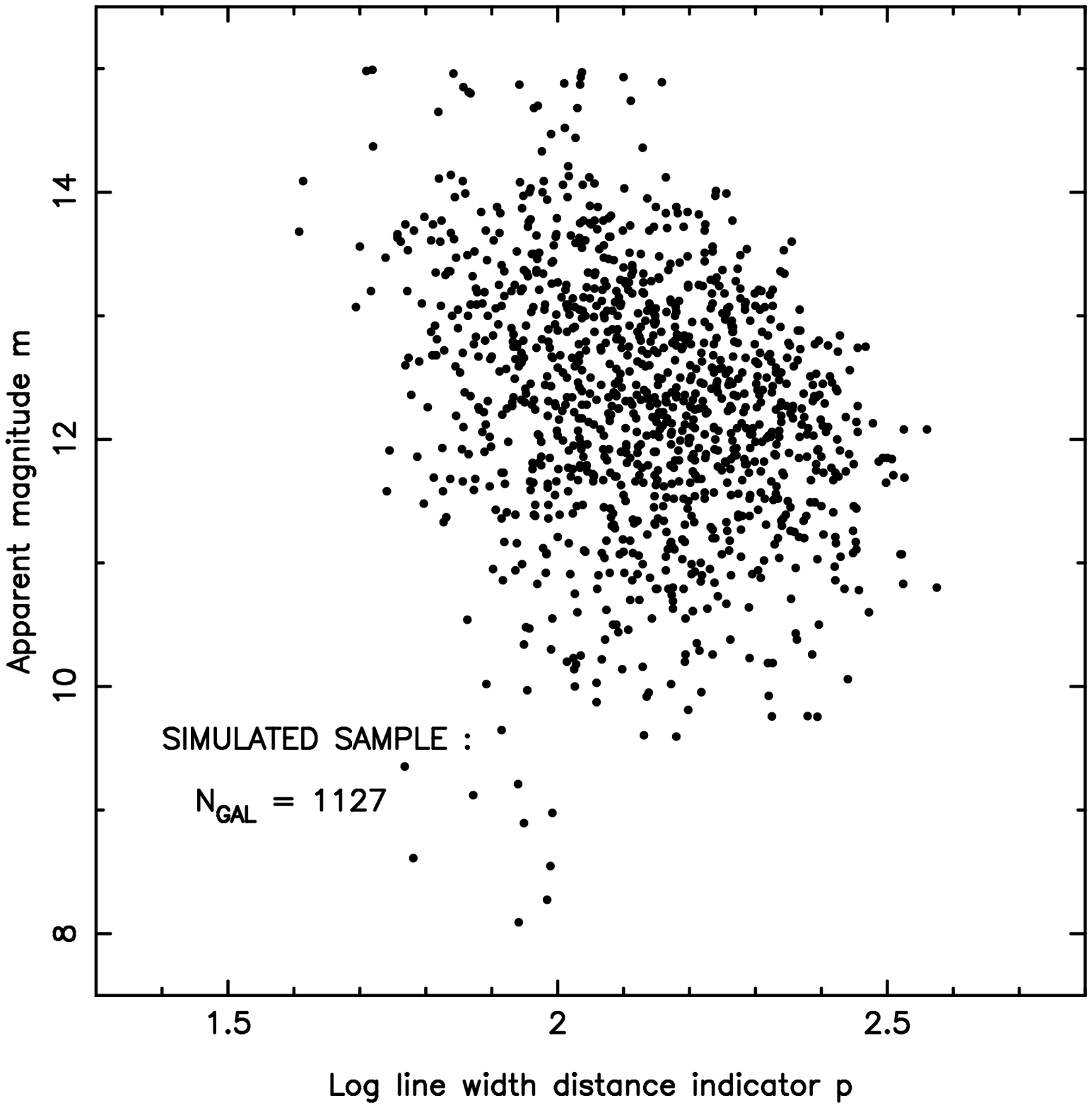,height=7.5 cm,width=8.7  cm,angle=270}
}}
{\hbox{
\psfig{file=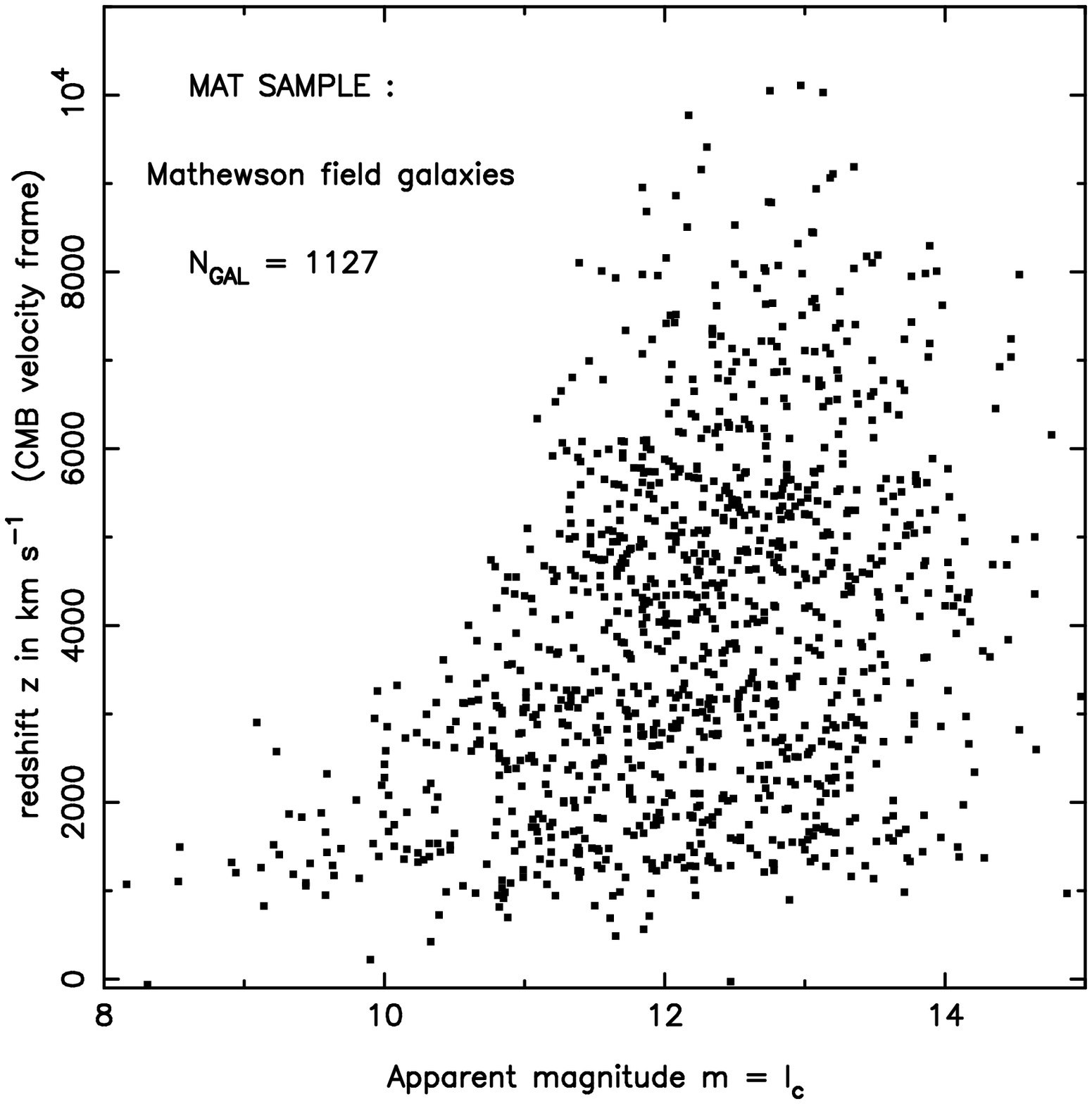,height=7.5 cm,width=8.7 cm,angle=270}
\psfig{file=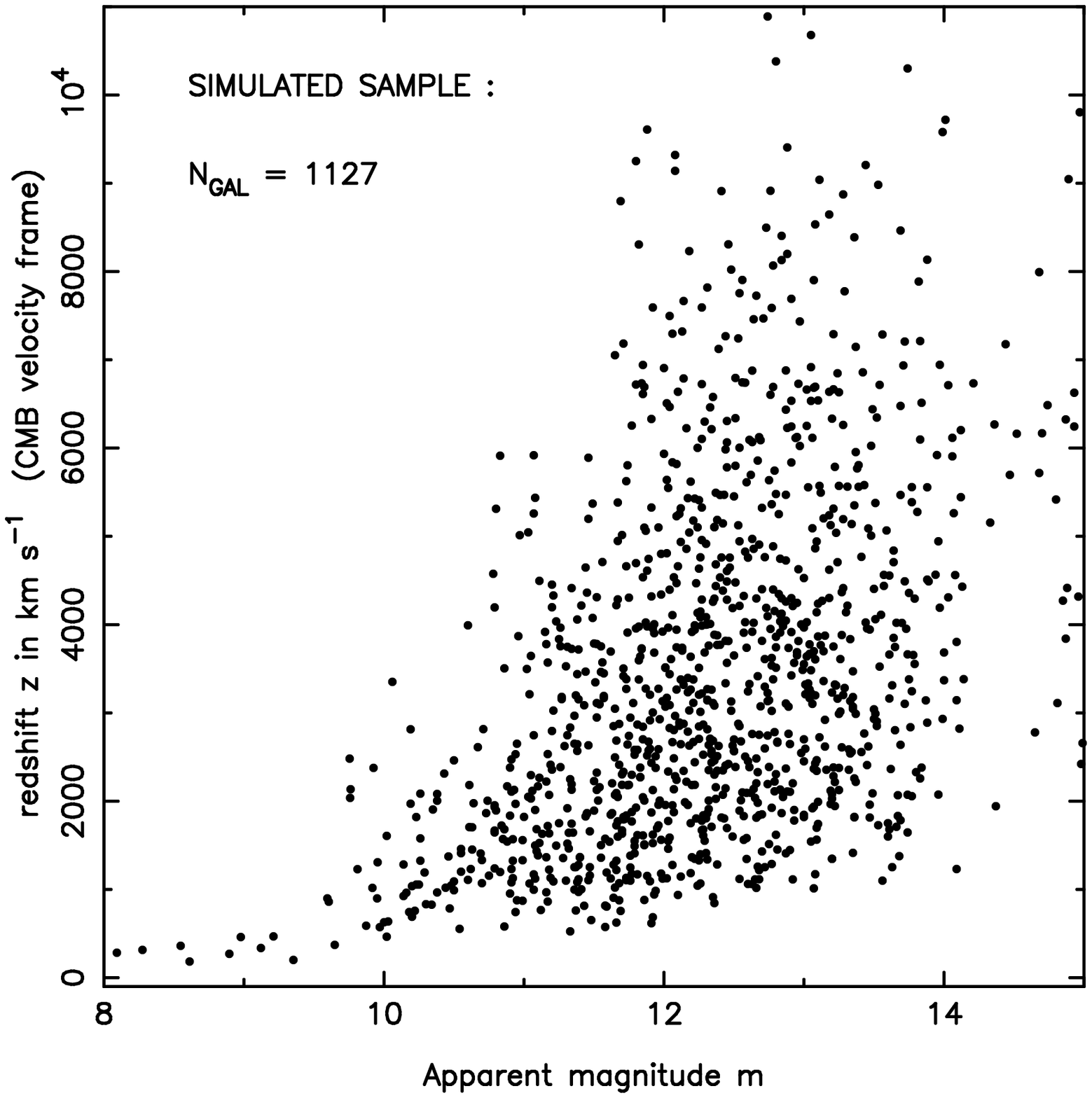,height=7.5 cm,width=8.7 cm,angle=270}
}}
\caption[]{
Mathewson spiral field galaxies sample versus simulated sample.}
\end{figure*}
\twocolumn
\section{Application to the Mathewson field galaxies sample}
\label{Application}

The spiral field galaxies sample of Mathewson
et al. \cite{Mat92} is a composite sample of
$N_{\rm gal}=1127$ spiral galaxies lying in the field
(galaxies identified as cluster members have
been excluded of this catalog). It covers the south hemisphere
and extends in redshift up to $10\,000$ km s$^{-1}$ with 
an effective depth of about $4\,000$ km s$^{-1}$.
Selection effects in observation are not trivial to model
since sampled galaxies have been
firstly selected in apparent diameter before Mathewson et al. 
apply their own selection criteria (minimum limits in
inclination and velocity rotation). Some
galaxies inherited from others observational programs
have been also included in the sample.
The data used hereafter in the analysis are :
\\
- the apparent magnitude $m~=~I_t^c$ where $I_t^c$ is
the total I-band apparent magnitude corrected for internal
and external extinction and K-dimming (column (6), second row in
Mathewson et al. \cite{Mat92}).
\\
- the log line-width distance indicator $p~=~log_{10}\,V_{\rm rot}$
where
$V_{\rm rot}$ is the maximum velocity of rotation of
the spiral galaxy (column (9) in
Mathewson et al. \cite{Mat92}).   
\\
- the redshift $z~=~V_{\rm CMB}$ in km s$^{-1}$ unit expressed
in the CMB velocity frame (column (11), first row in
Mathewson et al. \cite{Mat92}).
\\\indent
The application is presented as follows.
In section
\ref{Accuracies} is explained how numerical simulations, used
throughout the analysis for quantifying the amplitude of errors bars
and velocity biases, are performed. Section
\ref{Mathewsoncalibration} is devoted to test on the calibration
parameters proposed  by Mathewson et al. \cite{Mat92}. Finally, NCA
calibration of the Mathewson field galaxies sample is performed
section \ref{NCAcalibration}.
   
\subsection{Accuracies, amplitude of biases and simulations}
\label{Accuracies}

Accuracies of the NCA calibration parameter estimators
are herein calculated by using numerical simulations.
Reliability of the values of such standard deviations 
is of course
closely related to the way simulated samples succeed in
reproducing the characteristics of the real sample
under consideration. A preliminar study of the Mathewson
field galaxies (MAT) sample characteristics is thus required.

On figure 4 (top left) is shown the decimal logarithm
of the cumulative count in function of the apparent magnitude
$m$ for the MAT sample. It turns out that
the completeness in
magnitude of the MAT sample is violated beyond $m_{\rm sup}=11.3$,
corresponding to about $1/5$ of the total number of sampled galaxies.
Observational selection effects are then more complex than
a mere cut-off $m_{\rm lim}$ in apparent magnitude. In order
to mimic the real selection effects in observation affecting
the MAT sample, simulations are built in two steps.

A virtual sample, complete in apparent magnitude up
to $m_{\rm lim}=15.$, is firstly 
generated assuming the following characteristics :
\\
- Variable $p$ is generated according to a gaussian distribution
function $f_p(p)=g_G(p;p_0,\sigma_p)$.
\\
- Variable $\zeta^D$, accounting for the intrinsic scatter of
the DTF relation,  is generated according
to a gaussian distribution function
$g_G(\zeta^D;0,\sigma_\zeta^D)$.         \\
- The absolute magnitude $M=a^Dp+b^D-\zeta^D$ is then formed, with
$a^D$ and $b^D$ the slope and the zero-point of the DTF relation.
\\
- Variable $\mu$ is generated according to an exponential
distribution function $h(\mu)=\exp (3 \alpha \mu)$ and such that
$\mu < m_{\rm lim} -
M$ (i.e. distances are thus uniformly distributed in space).
\\
- The redshift $z=H_0\,\exp [\alpha (\mu - 25) ]$ with $H_0$
the Hubble's constant in km s$^{-1}$ Mpc$^{-1}$ is finally formed
according to the pure Hubble's flow hypothesis.
\\
Adopted values for the parameters
$p_0$,
$\sigma_p$,
$\sigma_\zeta^D$,
$a^D$,
$b^D$ and
$H_0$,
as well as their corresponding "zero-points" $B$ and $B^*$
introduced Eqs. (\ref{velocityestimator},\ref{Betoile}),
are given table 1.

The next step is to extract from this virtual sample
a subsamble of $N_{\rm gal}=1127$ which has the same distribution in
$m$ and $p$ than the observed distribution of the MAT
sample. In effect this selection is achieved as follows.
The $m$-$p$
plane is divided in boxes $box(i)$ of equal size and
the number $n(i)$
of MAT
galaxies belonging to each box $box(i)$ is memorized.
Boxes $box(i)$ are afterwards filled with galaxies belonging to 
the virtual sample while the observed $n(i)$ are not reached.
This selection procedure ensures that the $m$-$p$ distribution
of the simulated samples is approximately identical
to the observed one. It corresponds to introduce a
complex selection function $\phi_{\rm MAT}(m,p)$ in $m$ and $p$
directly derived from the data.
\begin{table}
\begin{center}
\caption[]{
Adopted values for the parameters of the
simulations}
\begin{tabular}{|c|c|c|c|c|c||c|c|} \hline
& & & & & & & \\
$p_0$&
$\sigma_p$&
$\sigma_\zeta^D$&
$a^D$&
$b^D$&
$H_0$&
$B$&
$B^*$\\
\hline
& & & & & & & \\
$1.82$&
$0.2$&
$0.35$&
$-7.$&
$-5.3$&
$85$&
$0.0097$&
$0.0106$\\
\hline
\end{tabular}
\end{center}
\end{table}

\begin{figure*}
%\begin{figure*}[htbp]
%\picplace{20. cm}
{\hbox{
\psfig{file=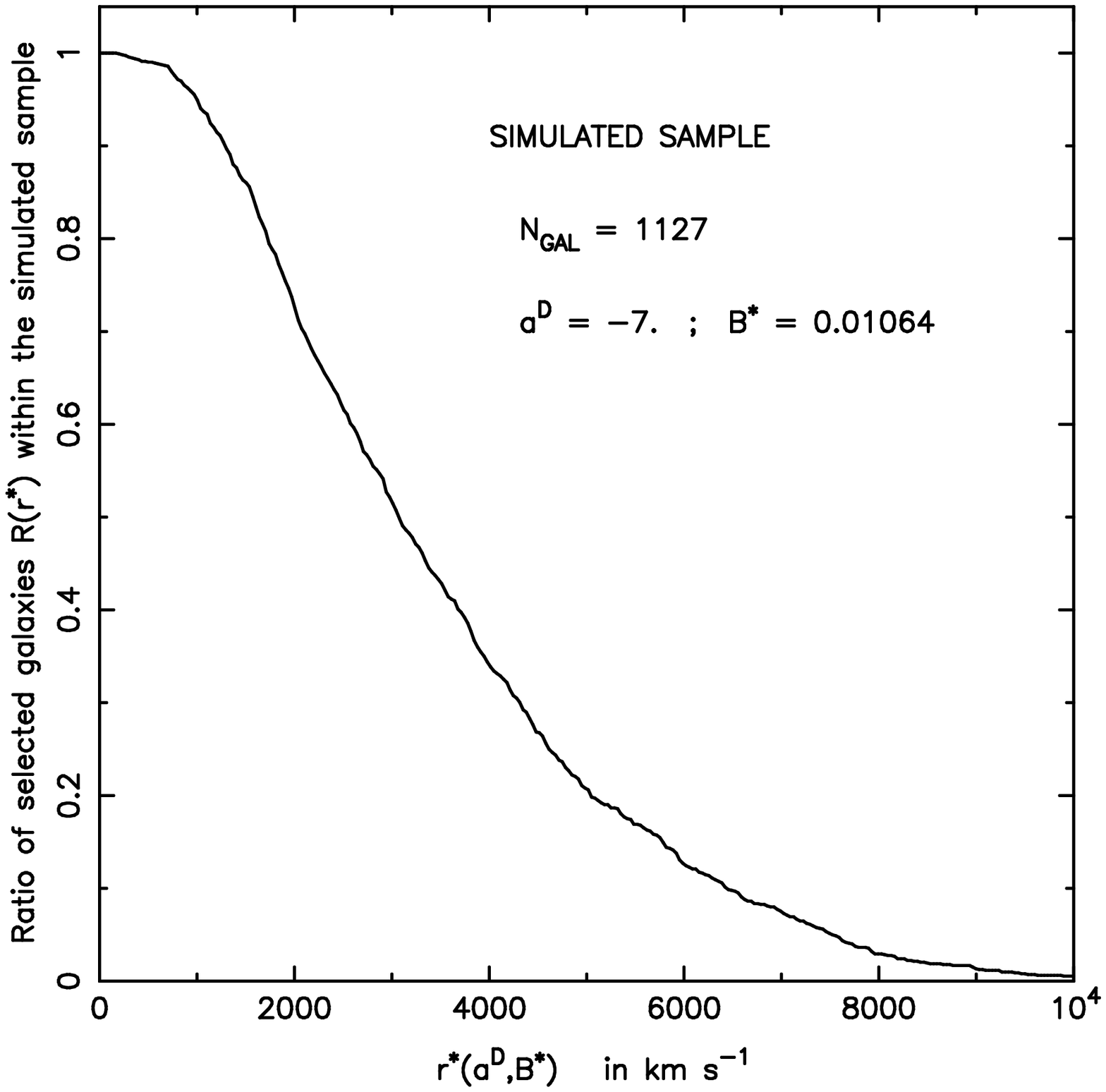,height=10. cm,width=9.2 cm,angle=270}
\psfig{file=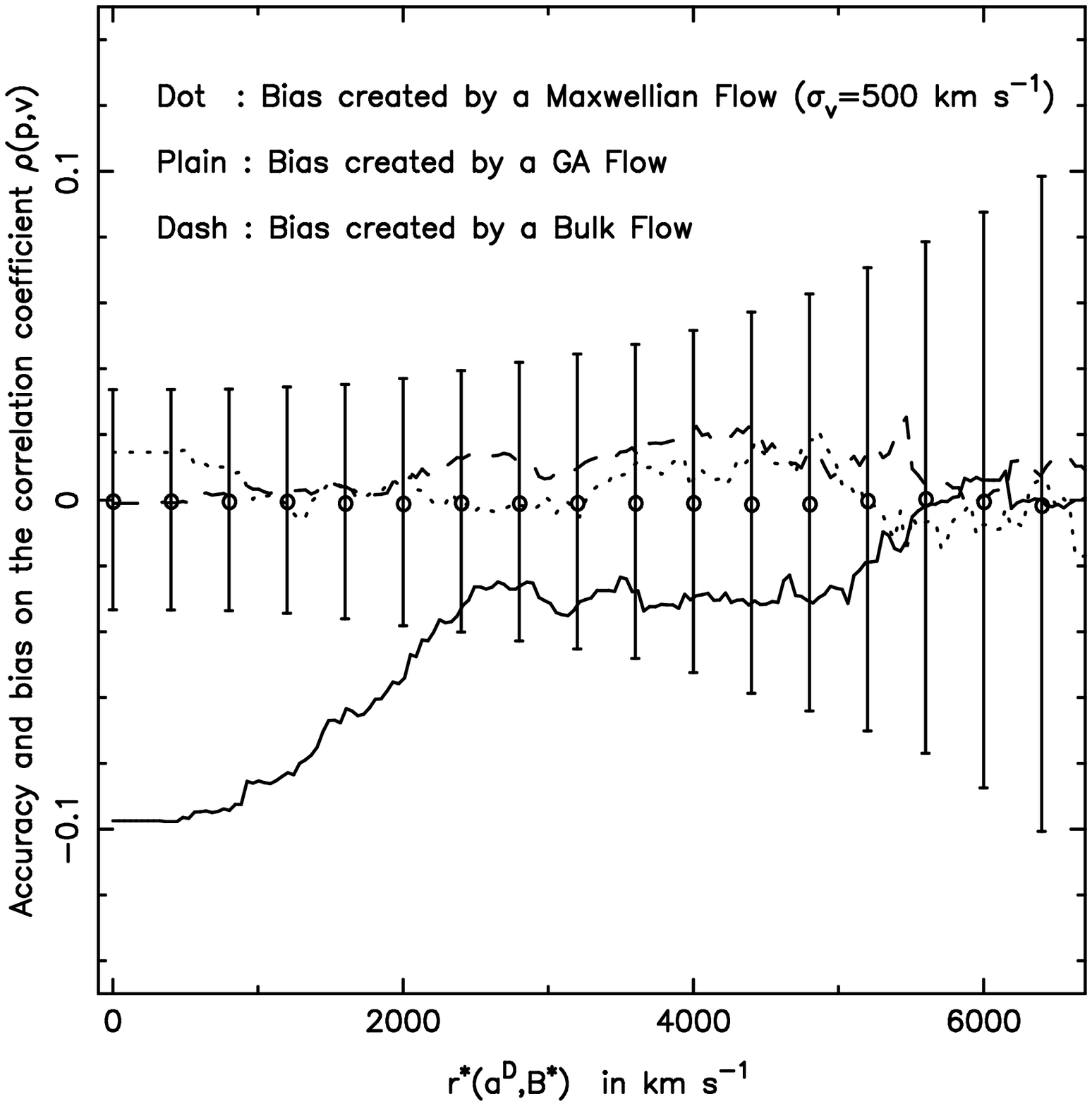,height=10. cm,width=9.2 cm,angle=270}
}}
{\hbox{
\psfig{file=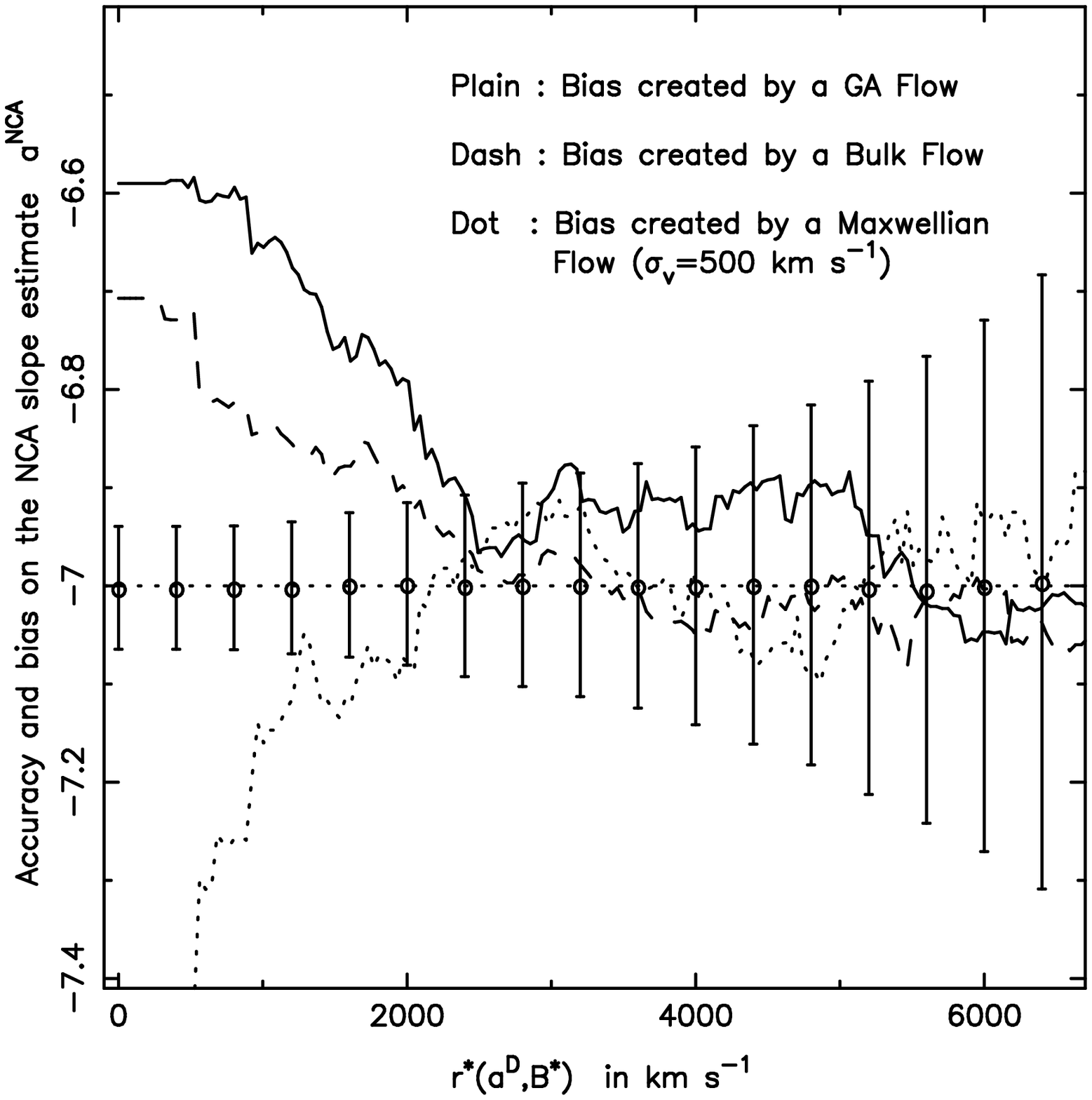,height=10. cm,width=9.2 cm,angle=270}
\psfig{file=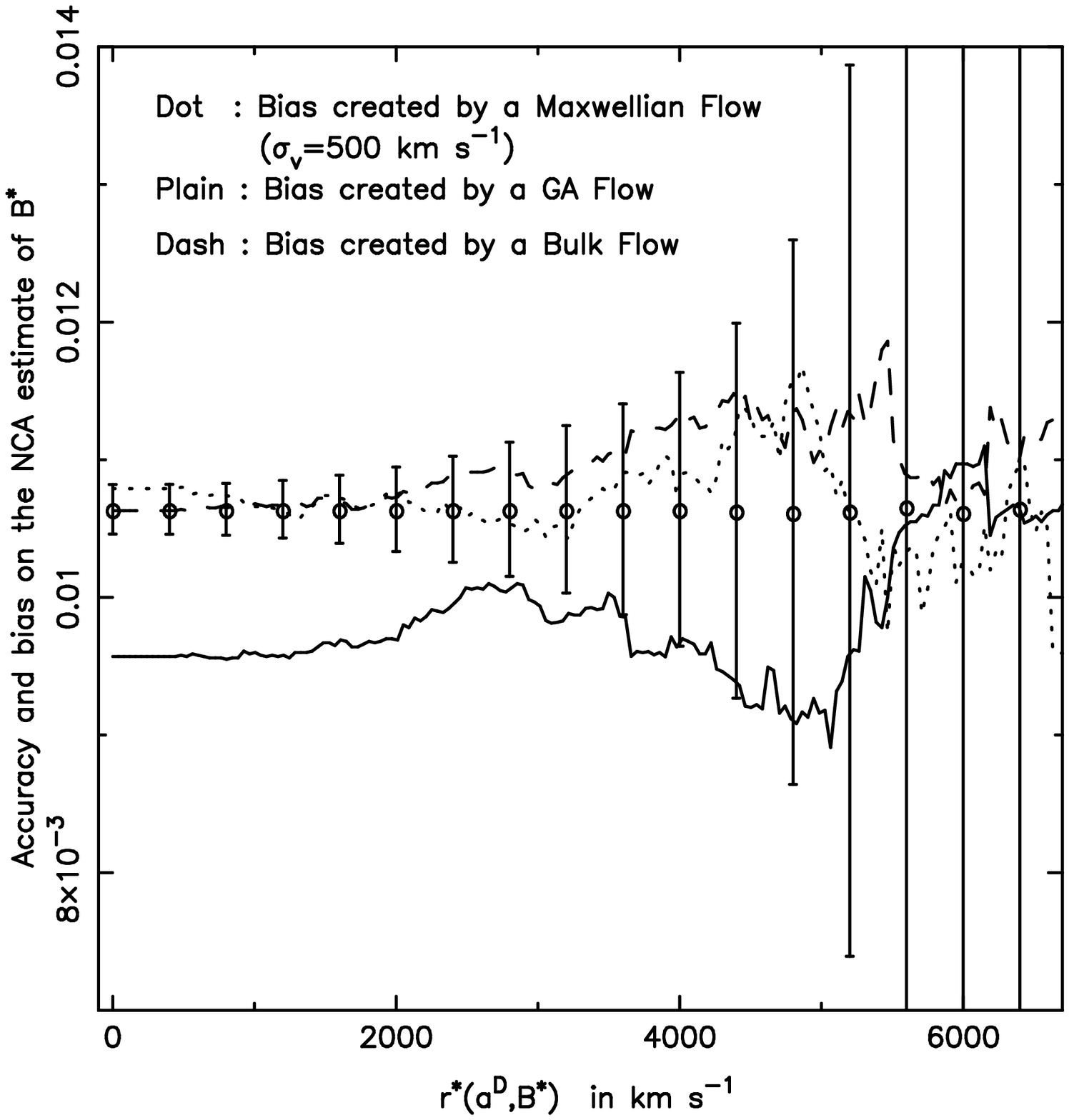,height=10. cm,width=9.2 cm,angle=270}
}}
\caption[]{
Accuracies of NCA estimators and amplitudes of velocity biases :
(Top left) Ratio $R(r^*)$ of selected galaxies within
the simulated
samples function of the extra cut-off in distance estimate
$r^*(a^D,B^*)$.
(Top right) Standard deviation for the correlation
coefficient $\rho(p,{\tilde v})$ and velocity biases created by GA,
Bulk and Maxwellian flows.
(Bottom left) Standard deviation for NCA slope estimator 
$a^{\rm NCA}$ and velocity biases created by GA, Bulk and
Maxwellian flows.
(Bottom right) Standard deviation for NCA "zero-point" estimator 
$B_*^{\rm NCA}$ and velocity biases created by GA, Bulk and
Maxwellian flows.
}
\end{figure*}

Figure 4 (top right) shows the cumulative count in apparent
magnitude for the MAT sample and for a simulated sample.
The cumulative count expected
for a sample complete up to $m_{\rm lim}=15.$
is also shown for comparison. Figures 4 (center left) and
(center
right) show respectively the distribution in the $m$-$p$ plane
of the MAT sample and a simulated sample. The difference
between these two distributions are due to discretization
effects which appear when applying the boxes algorithm.
Figures 4 (bottom left) and (bottom right) show respectively
the $m$-$z$ distributions for the MAT sample and for
a simulated sample. The fact that simulated sample
distribution approximately reproduces the observed one
is encouraging. It means that the working hypotheses assumed
when generating simulated samples, such as the uniform spatial
distribution of galaxies, are close to be verified by
the Mathewson field galaxies sample.
The likeness to data can certainly be improved,
by choosing a more realistic shape for the luminosity
function for example (i.e. the $p$ distibution $f_p(p)$\,),
but is out of the scope of this study. Hereafter, these
simulated samples
will be considered as fair representatives of
the observed catalog.

Figure 5 visualizes results obtained on these simulated samples.
At the top left is plotted the ratio $R(r^*)$ of selected galaxies
within a simulated sample when the subsampling in distance
estimate (i.e. ${\tilde r}\,\,>\,\,r^*(a^D,B^*)$\,) is applied.
Standard statistical deviation of the correlation coefficient
between $p$ and the velocity estimates ${\tilde v}(a^D,B^*)$
in function of the cut-off in distance estimate $r^*$
is shown figure 5 (top right). Standard deviation
(i.e. accuracy) of 
the NCA slope estimator
$a^{\rm NCA}$ and of the NCA "zero-point" $B_*^{\rm NCA}$ are
respectively presented figures 5 (bottom left) and (bottom right).
If no peculiar velocity field is present (i.e. pure Hubble
flow hypothesis, as it is the case for simulated samples), we see
that $a^{\rm NCA}$ and $B_*^{\rm NCA}$ estimators are not biased, as it was
previously proven appendices B and C (averaged over $1000$
simulations, their values coincide with the input slope and
"zero-point" of the simulated sample). It illustrates one of the
potentialities of the null-correlation approach for calibrating TF
like relations, i.e. its insentivity to observational selection
effects in apparent magnitude $m$ and log line-width distance
indicator $p$.
For these simulated samples supposed to mimic the
Mathewson field
galaxies sample ($N_{\rm gal}=1127$), accuracy
$\sigma_a$ of the NCA slope estimator $a^{\rm NCA}$ 
sounds clearly good : $\sigma_a=0.07$
or $1\%$  of $a^D$ if all the sample is selected, $\sigma_a=0.1$ 
if half of the nearby galaxies of
the sample are discarded (i.e. $R(r^*)=0.5$ 
or $r^* \approx 3000$ km s$^{-1}$)
and $\sigma_a=0.15$ at $r^* \approx 4000$ km s$^{-1}$
(i.e. $R(r^*)=0.25$) \footnote{The standard deviation $\sigma_a$
is proportional to $R(r^*)^{-1/2}$, as expected.}.
The same  remark holds for the NCA "zero-point" $B_*^{\rm NCA}$ accuracy
$\sigma_{B^*}$, $\sigma_{B^*}=0.0003$ or $3\%$  of $B^*$
at $r^*=0$ and $\sigma_{B^*}=0.001$ at $r^* \approx 4000$ km
s$^{-1}$. 

Influence of peculiar velocity field on the NCA estimators
is analysed using three examples : "Great Attractor" (GA) flow,
constant or bulk flow and gaussian random or Maxwellian flow.
In order to take into account the peculiarity
of the 3D spatial distribution of the Mathewson field galaxies sample,
biases created by these flows have
been calculated by comparing the estimates of $a^{\rm NCA}$, $B_*^{\rm NCA}$ and
$\rho(p,{\tilde v})$ when one of these velocity field is added
to the observed redshifts of the MAT sample, with these
estimates for the real MAT sample. The Maxwellian flow
has a velocity agitation of $\sigma_v=500$ km s$^{-1}$ and
the bulk flow has been chosen to point toward direction
$l=310$ and $b=20$ in galactic coordinates with an amplitude
of $500$ km s$^{-1}$. The GA
flow \footnote{Since the real distance of MAT galaxies are not known, the
contribution of the radial peculiar velocity inferred
by GA flow model to observed redshift has been added
under the approximation that galaxies distances are given by their
observed redshifts.}   is the one of Bertschinger et al.
\cite{Ber88}, centered at a distance of $4\,200$ km s$^{-1}$ toward
$l=310$ and $b=20$ and creating an infall velocity for our Local
group of $535$ km s$^{-1}$. 

Figure 5 (bottom left) illustrates particularly well
the discussion of section \ref{slopecalibration} which
was based on analytical results.
Biases on the NCA slope estimate
$a^{\rm NCA}$ created by the presence of
Maxwellian and Bulk flows become negligible when
nearby galaxies of the MAT sample are discarded
using the subsampling procedure in distance estimate
(say herein for $r^* > 3000$ km s$^{-1}$).
Influence of GA flow can be controled likewise.
In practice, large values of the cut-off in distance
estimate $r^*$ have to be preferred, of course with
regards to the accuracy and to the $r^*$-dependent statistical
fluctuations affecting the NCA slope estimate at this
distance $r^*$.  

Influences of velocity fields on $B_*^{\rm NCA}$ and
$\rho(p,{\tilde v})$ are shown respectively figures 5 (bottom
right) and (top right). As expected,
Maxwellian flow does not bias these two quantities. Since
the calibration of the simulated samples was not  
performed line-of-sight by line-of-sight, presence
of Bulk flow slightly biases $B_*^{\rm NCA}$ and
$\rho(p,{\tilde v})$. On the other hand, we can
see that GA flow creates a significative bias
on the two quantities (i.e. greater than their standard
deviations). The fact that these biases vanish
for large values of the cut-off in distance estimate $r^*$
is due to the specific form of the GA flow and to
the characteristics of the 3D spatial distribution of
the MAT sample. It cannot be interpreted as a general
feature since the correlation between $p$ and ${\tilde v}$
is not expected to vanish when the subsampling in
distance estimate is applied, as it is the case
for the $p$-$X(a^D)$ correlation. Some reasons
may however be advocated for favouring  large
values of the cut-off in distance estimate $r^*$.
Since such a subsampling selects preferencially far away
galaxies, a slighter coherence of their peculiar velocities is
expected, consequently to their mutual distances.  
Finally, one remarks
that amplitude of the bias created by huge flows such
as the "Great Attractor" is not greater than
$\Delta B^*\approx 0.001$, or $10\%$ of the value of $B^*$.
Same remark for the bias on the correlation coefficient
$\rho(p,{\tilde v})$ which is less than $\Delta \rho=0.1$
whatever the value of the cut-off in distance estimate $r^*$.     

\subsection{Testing on Mathewson et al. calibration
parameters} \label{Mathewsoncalibration} 

\begin{figure*}
%\begin{figure*}[htbp]
%\picplace{9.5 cm}
{\hbox{
\psfig{file=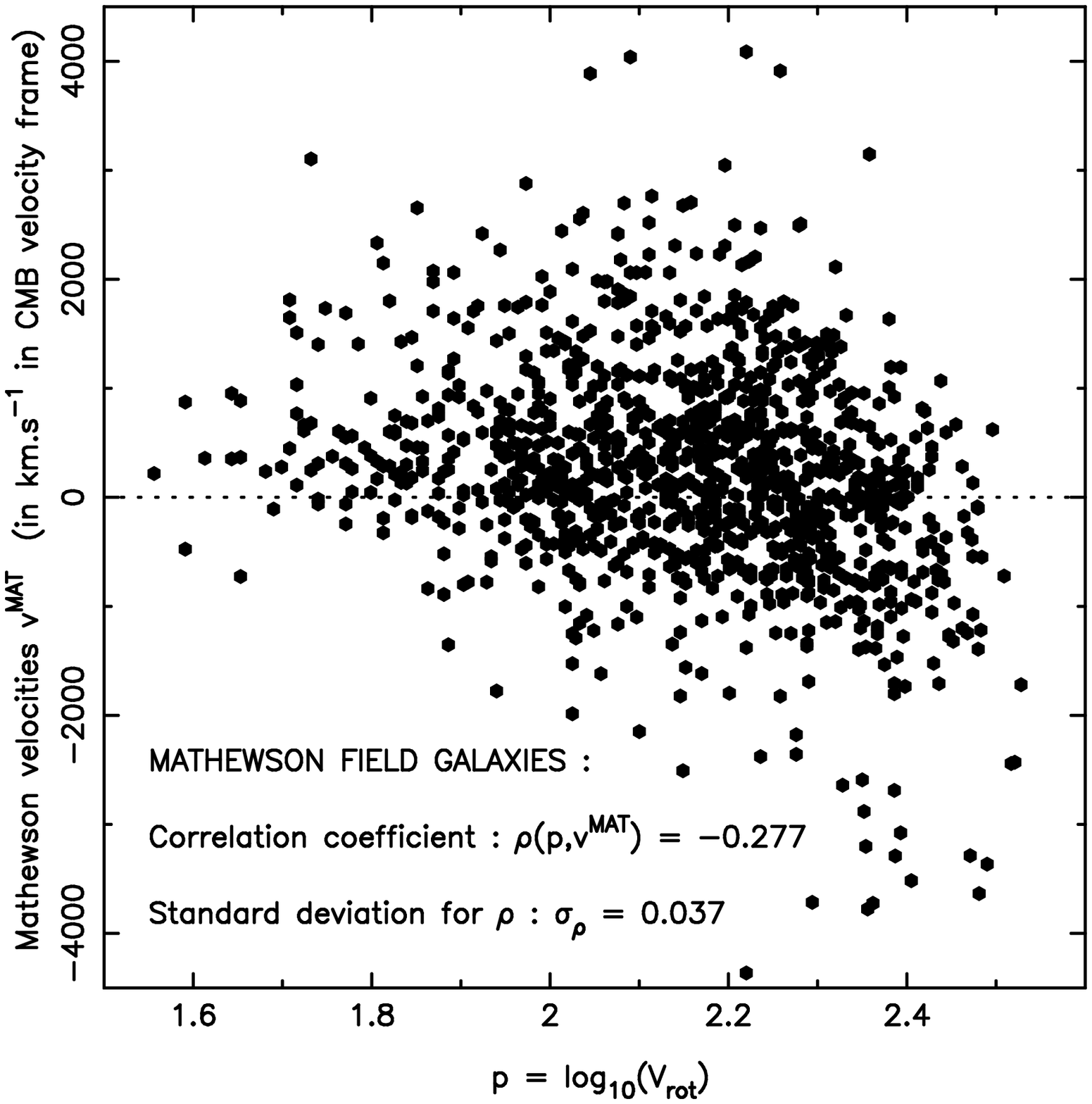,height=9.5 cm,width=9.4 cm,angle=270}
\psfig{file=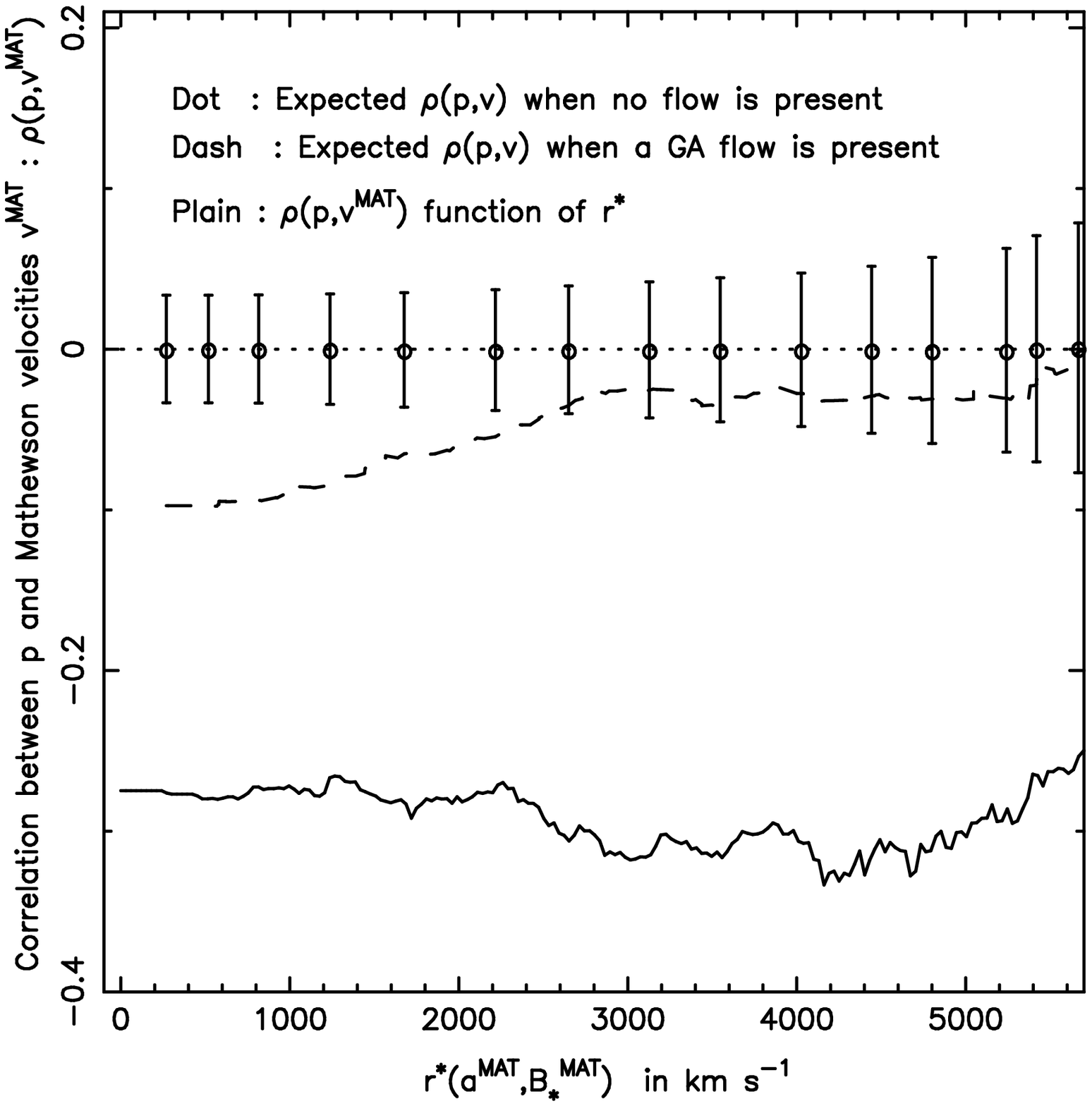,height=9.5 cm,width=9.  cm,angle=270}
}}
\caption[]{
Correlation between $p$ and Mathewson velocity estimates 
${\tilde v}^{\rm MAT}$.
}
\end{figure*}

Mathewson et al. \cite{Mat92} have proposed some values
for the DTF calibration parameters. The authors
calibrate the DTF slope $a^D$ in the Fornax cluster
($14$ galaxies). An estimate of $a^{\rm MAT}=-7.96$ is
obtained by performing a
linear regression in magnitude. Assuming that the
true distance in km s$^{-1}$ units of Fornax cluster
is $1\,340$ km s$^{-1}$, authors derived a value
of $B^{\rm MAT}=0.0039$ for DTF relative zero-point $B$.
If $H_0=85$
km s$^{-1}$ Mpc$^{-1}$, it implies a value of $b^{\rm MAT}=-3.31$ for
the DTF zero-point $b^D$. Averaging over the seven richest clusters
of their catalog, a value of $\sigma_\zeta^{\rm MAT}=0.32$
is estimated for the intrinsic scatter of the DTF relation,
which gives a value of $B_*^{\rm MAT}=0.0042$ for the
DTF relative "zero-point" $B^*$. No error bars are
proposed for these estimates.

Mathewson et al. DTF calibration parameters are tested on
figure 6. Figure 6 (left) shows the correlation between
$p$ and Mathewson et al. peculiar velocity estimate
${\tilde v}^{\rm MAT}$ for the $N_{\rm gal}=1127$ of
the MAT sample. Figure 6 (right) shows variations
of $\rho(p,{\tilde v}^{\rm MAT})$ with respect to
the cut-off in distance estimate $r^*$ and the deviations
from 0 expected when a flow such as the "Great Attractor"
is present. It looks very unlikely that the strong
correlation found between $p$ and ${\tilde v}^{\rm MAT}$
can be explained by the presence of large scale coherent
peculiar velocity field. The same feature is observed
for the $p$-$X(a^{\rm MAT})$ correlation (not shown).
This test leads to question in the validity of the calibration
techniques used by Mathewson et al. when deriving
slope and relative zero-point of the DTF relation.

As a matter of fact, it is known that the estimator
of the DTF slope $a^D$, obtained in a cluster
by a linear regression on $m$, is biased by the presence
of observational selection effects on $m$ and $p$
(see Lynden-Bell et al. \cite{Lyn88} for example).
This bias can be corrected on in theory but a full
description of selection effects in observation is
thus required (see for example Willick \cite{Wil94}), 
and so is difficult to realize in practice.
Same kind of remark holds for the relative zero-point
estimate $B$. Since existing peculiar velocity field models
proposed in the literature are far to be perfect nor
accurate, a room of uncertainty remains when attributing
a true distance to the calibration cluster (an error
of $250$ km s$^{-1}$ for the assumed true distance of Fornax
cluster will imply a relative error of $\frac{\delta B}{B}
\approx 20\%$ for the relative zero-point $B$).

\subsection{NCA calibration of the Mathewson spiral
field galaxies sample} \label{NCAcalibration} 
\begin{figure*}
%\begin{figure*}[htbp]
%\picplace{20. cm}
{\hbox{
\psfig{file=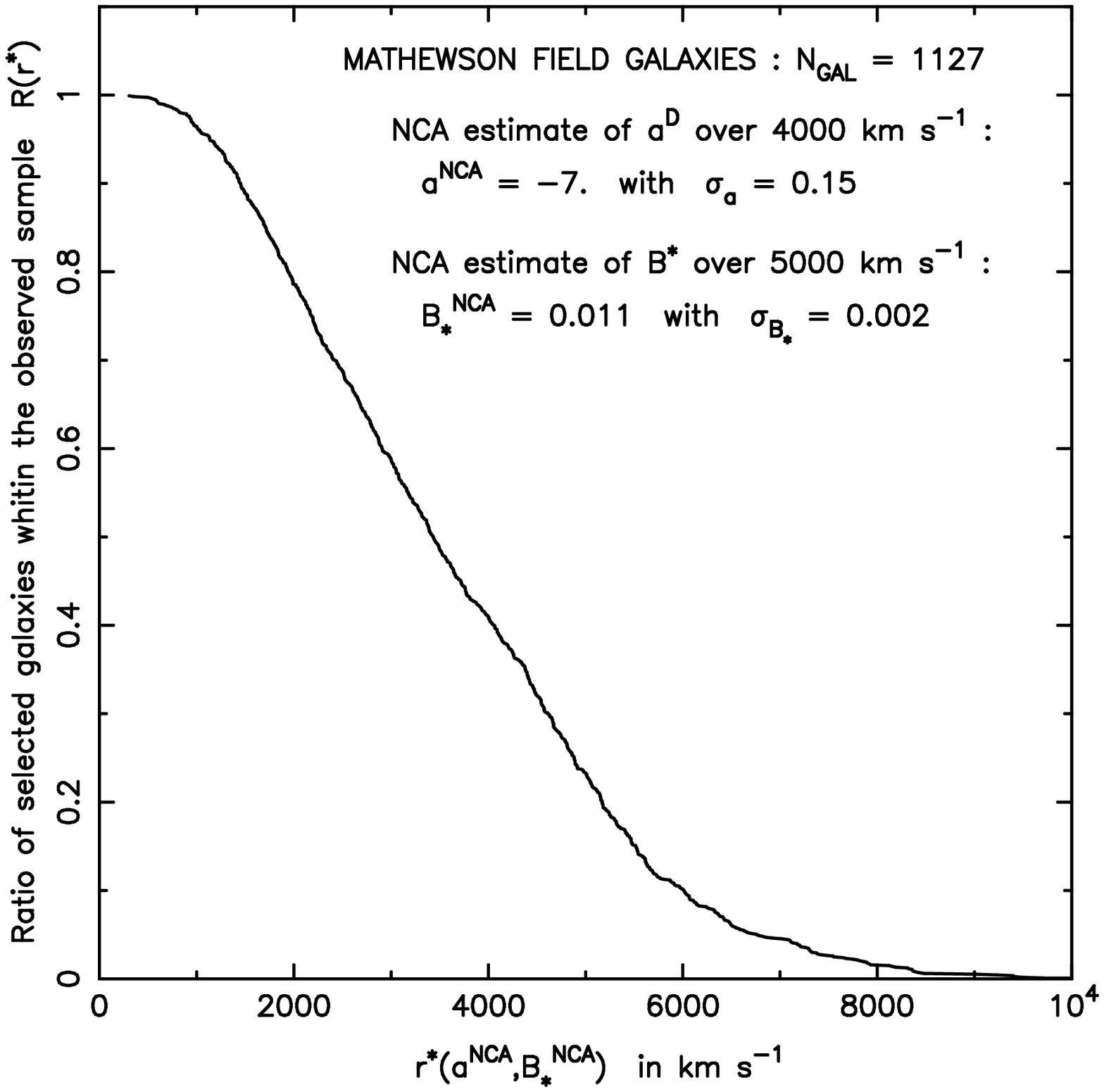,height=10. cm,width=9.2 cm,angle=270}
\psfig{file=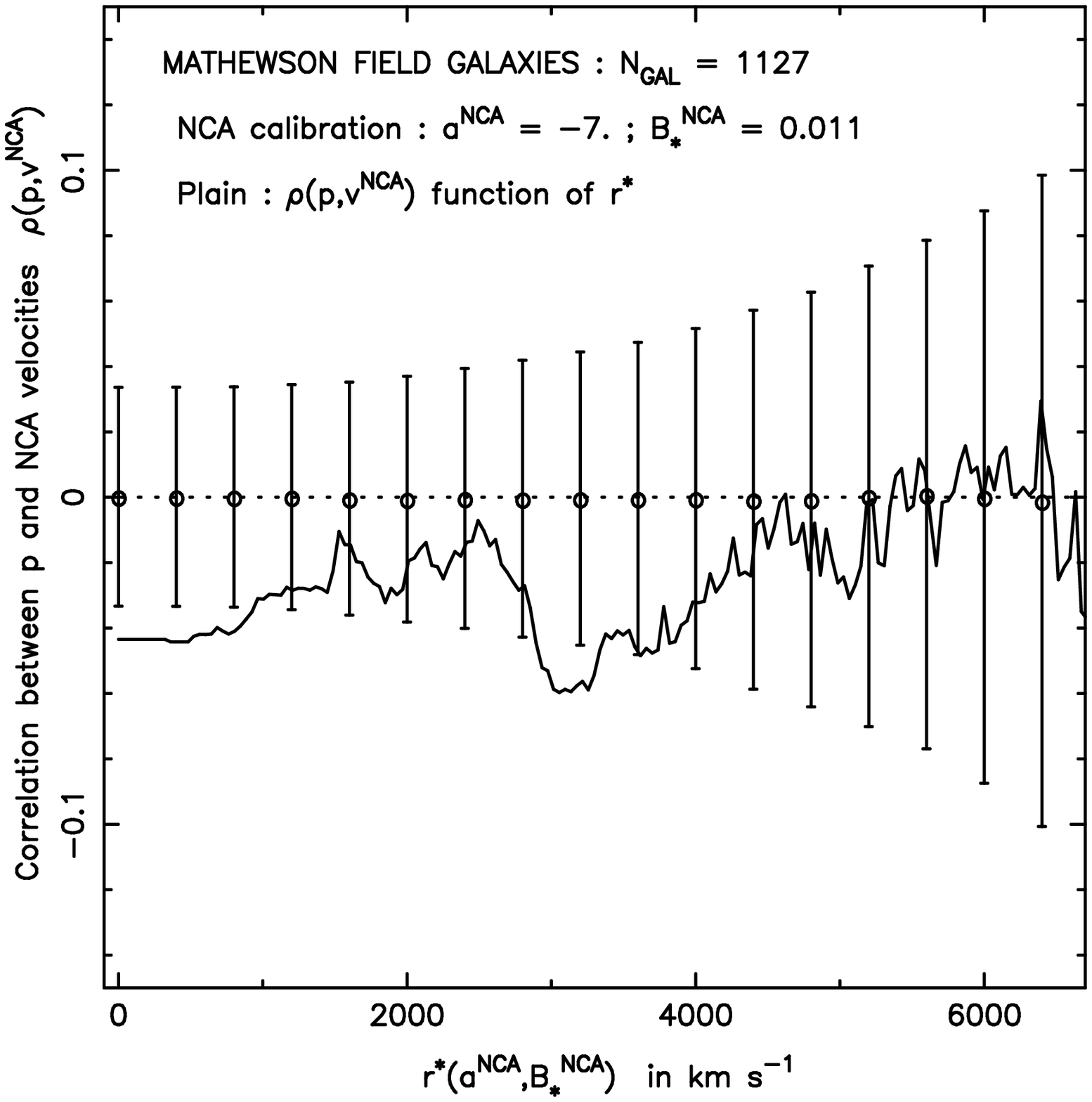,height=10. cm,width=9.2 cm,angle=270}
}}
{\hbox{
\psfig{file=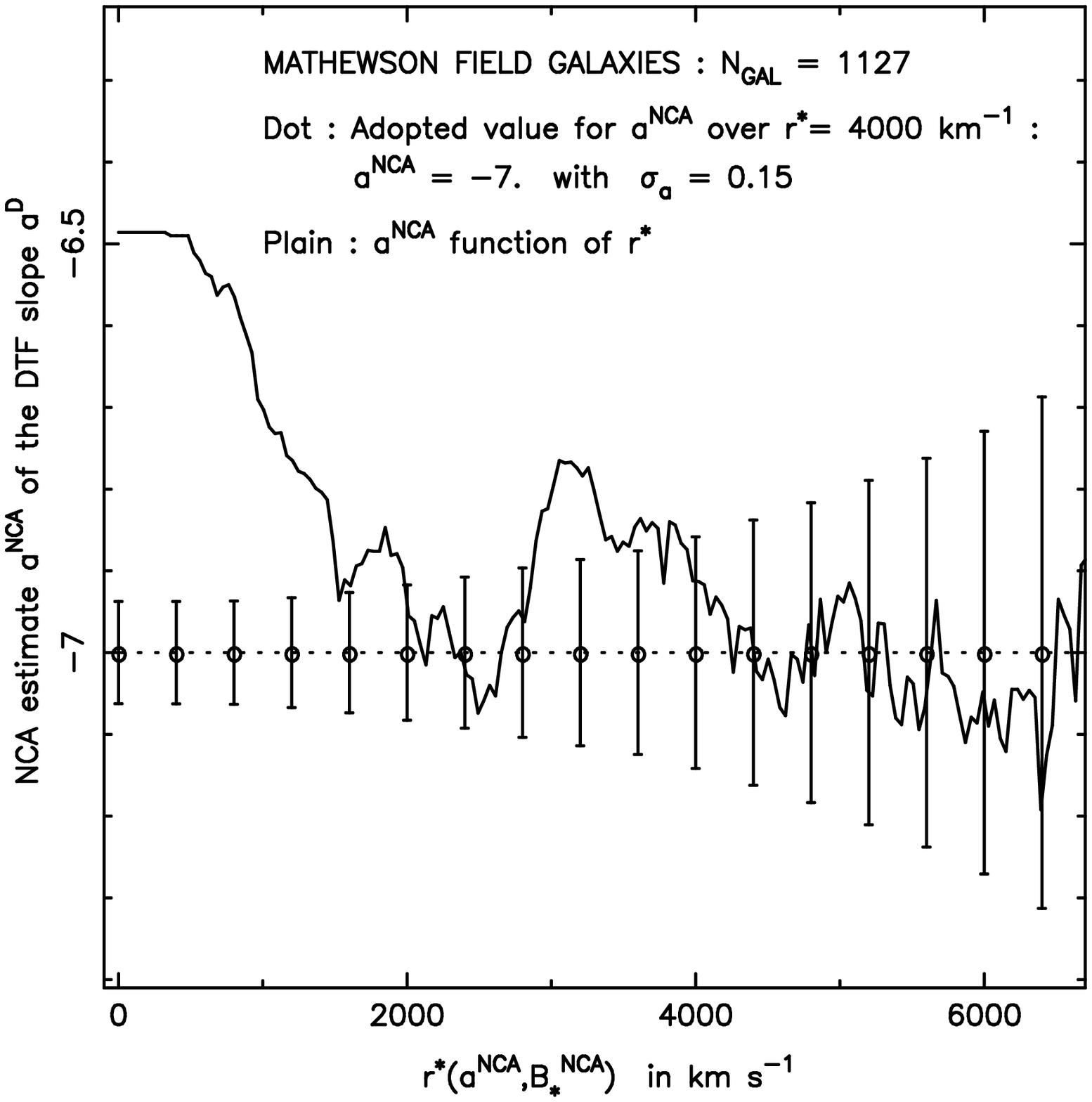,height=10. cm,width=9.2 cm,angle=270}
\psfig{file=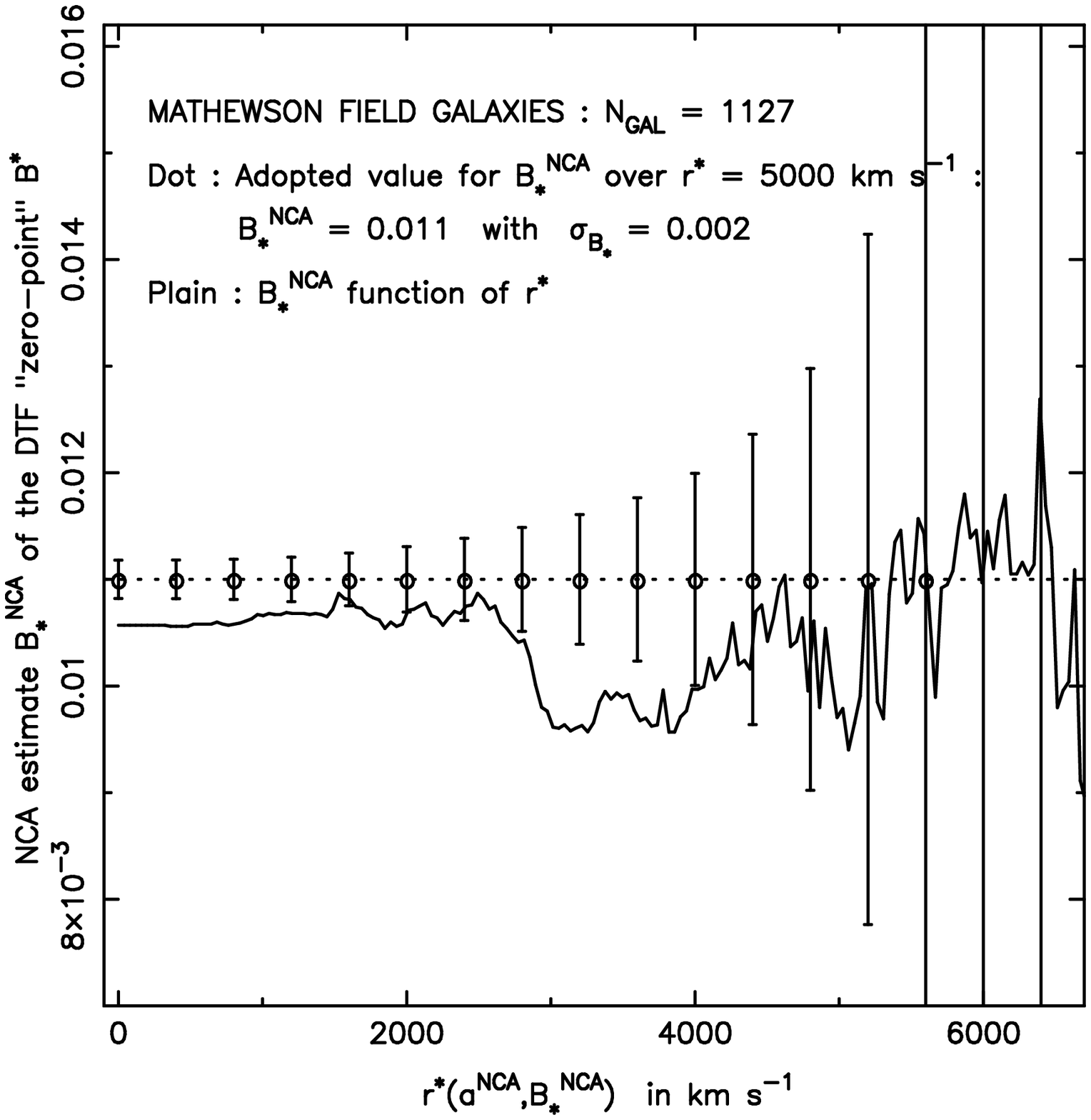,height=10. cm,width=9.2 cm,angle=270}
}}
\caption[]{
NCA calibration of the Mathewson spiral field galaxies sample :
(Top left) Ratio $R(r^*)$ of selected galaxies within
the MAT sample function of the extra cut-off in distance estimate
$r^*(a^{\rm NCA},B_*^{\rm NCA})$.
(Top right) Correlation
coefficient $\rho(p,{\tilde v}^{\rm NCA})$ between $p$ and
NCA peculiar velocity estimate ${\tilde v}^{\rm NCA}$.
(Bottom left) NCA slope estimate 
$a^{\rm NCA}$ function of the extra cut-off in distance estimate
$r^*(a^{\rm NCA},B_*^{\rm NCA})$.
(Bottom right) NCA relative "zero-point" estimate 
$B_*^{\rm NCA}$ function of the extra cut-off in distance estimate
$r^*(a^{\rm NCA},B_*^{\rm NCA})$.
}
\end{figure*}

\begin{figure*}
%\begin{figure*}[htbp]
%\picplace{9.5 cm}
{\hbox{
\psfig{file=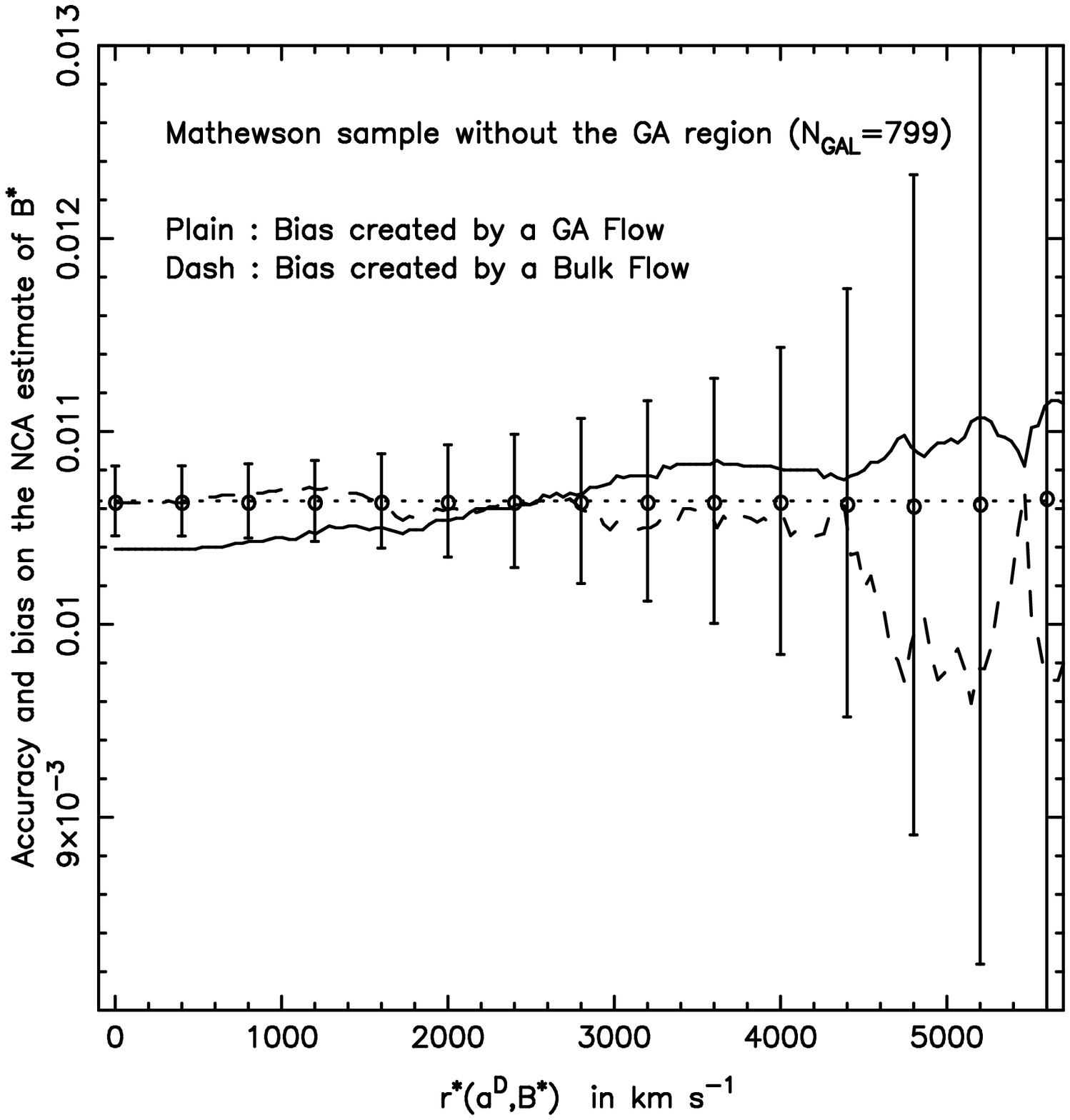,height=9.5 cm,width=9.2 cm,angle=270}
\psfig{file=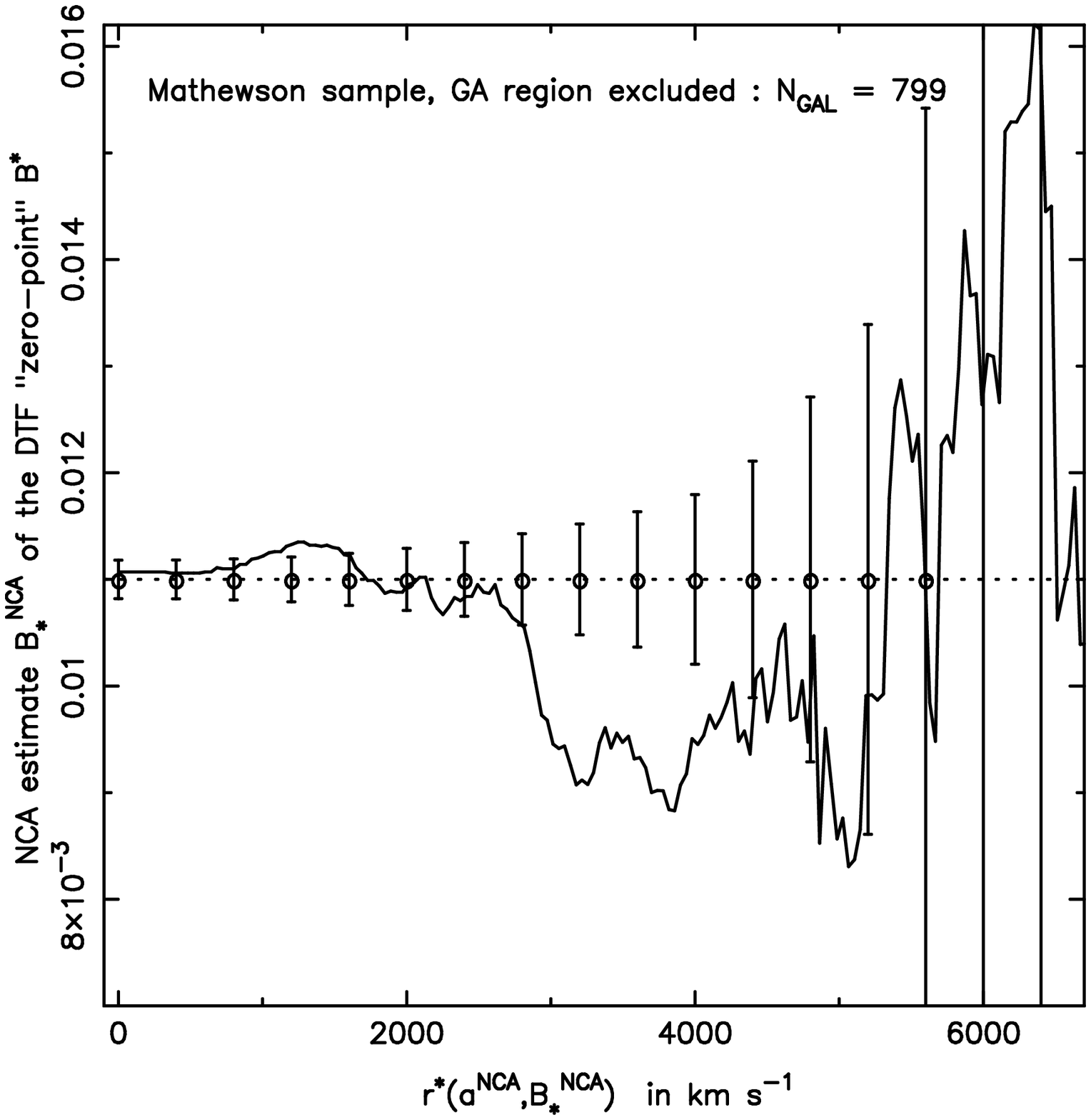,height=9.5 cm,width=9.2 cm,angle=270}
}}
\caption[]{
NCA "zero-point" estimate for $N_{\rm gal}=799$ remaining
galaxies of the MAT sample when excluding the GA region
(defined as a conic area pointing toward $l=310$ and
$b=20$ with an angular aperture of $45^{\rm o}$) :
(Left) Standard deviation for NCA "zero-point" estimator 
$B_*^{\rm NCA}$ and velocity biases created by GA and Bulk flows.
(Right) NCA "zero-point" estimate 
$B_*^{\rm NCA}$ function of the extra cut-off in distance estimate
$r^*(a^{\rm NCA},B_*^{\rm NCA})$.
}
\end{figure*}
\begin{figure*}
%\begin{figure*}[htbp]
%\picplace{20.5 cm}
{\hbox{
\psfig{file=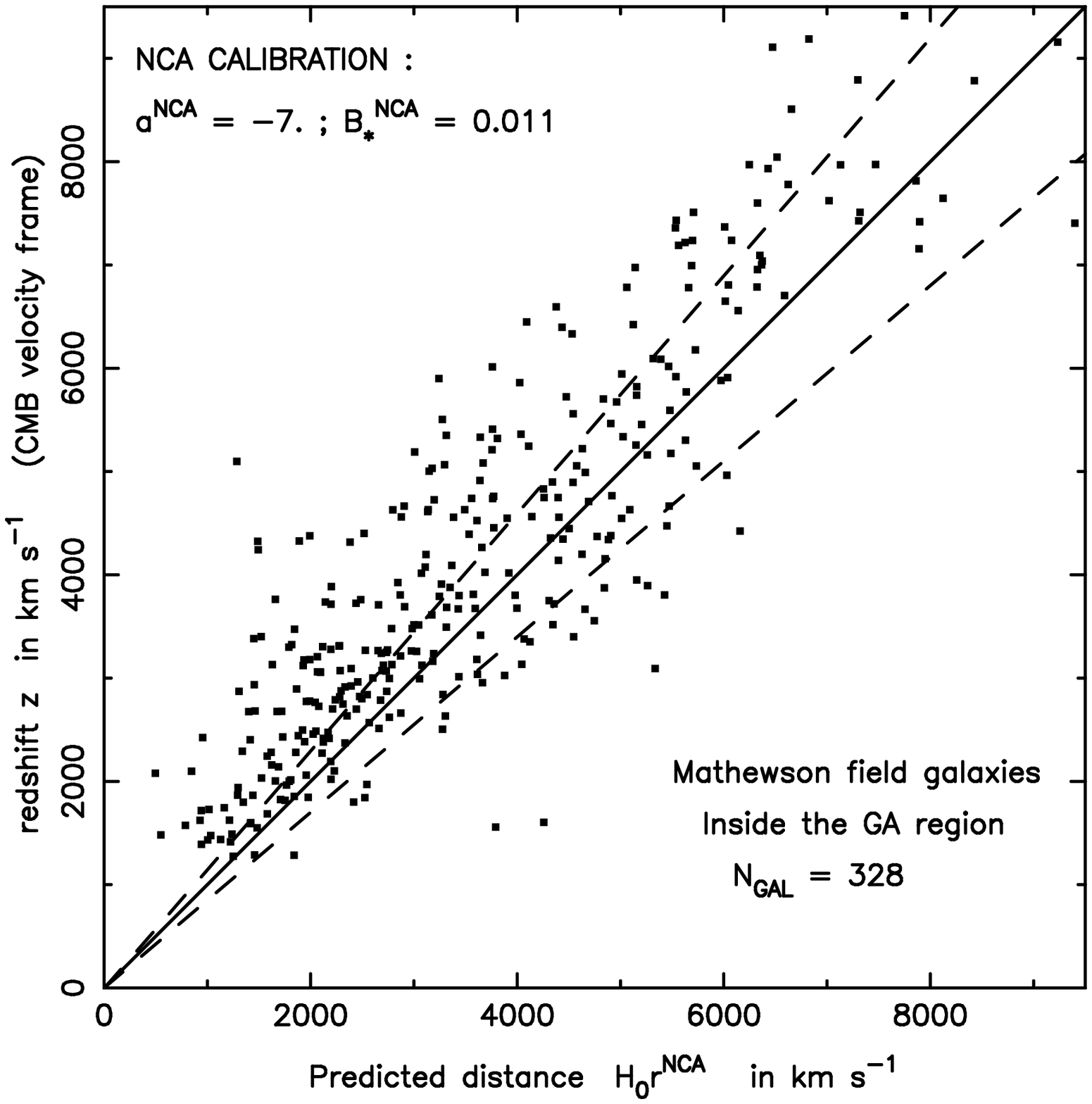,height=10. cm,width=9.2 cm,angle=270}
\psfig{file=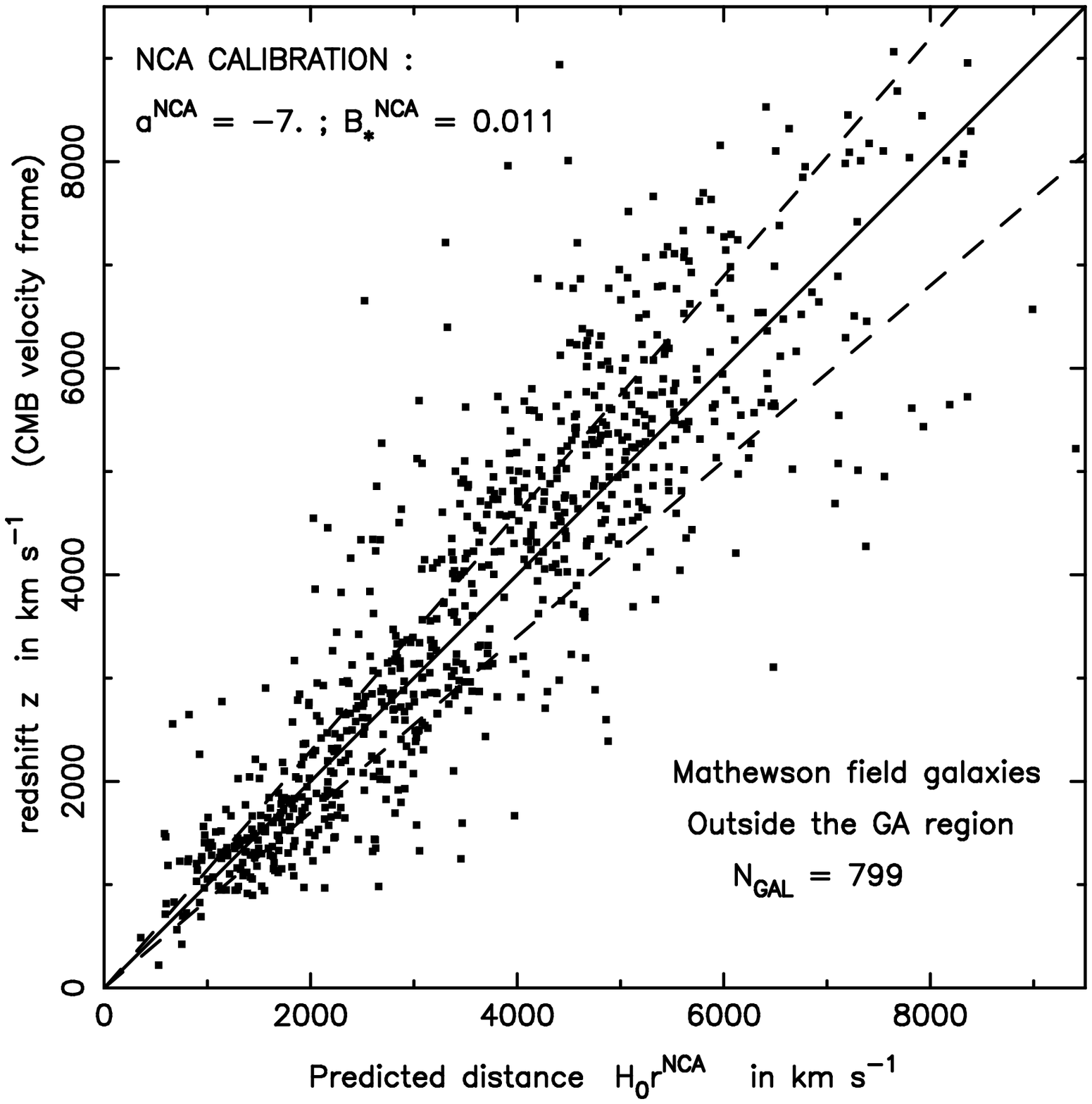,height=10. cm,width=9.2 cm,angle=270}
}}
\psfig{file=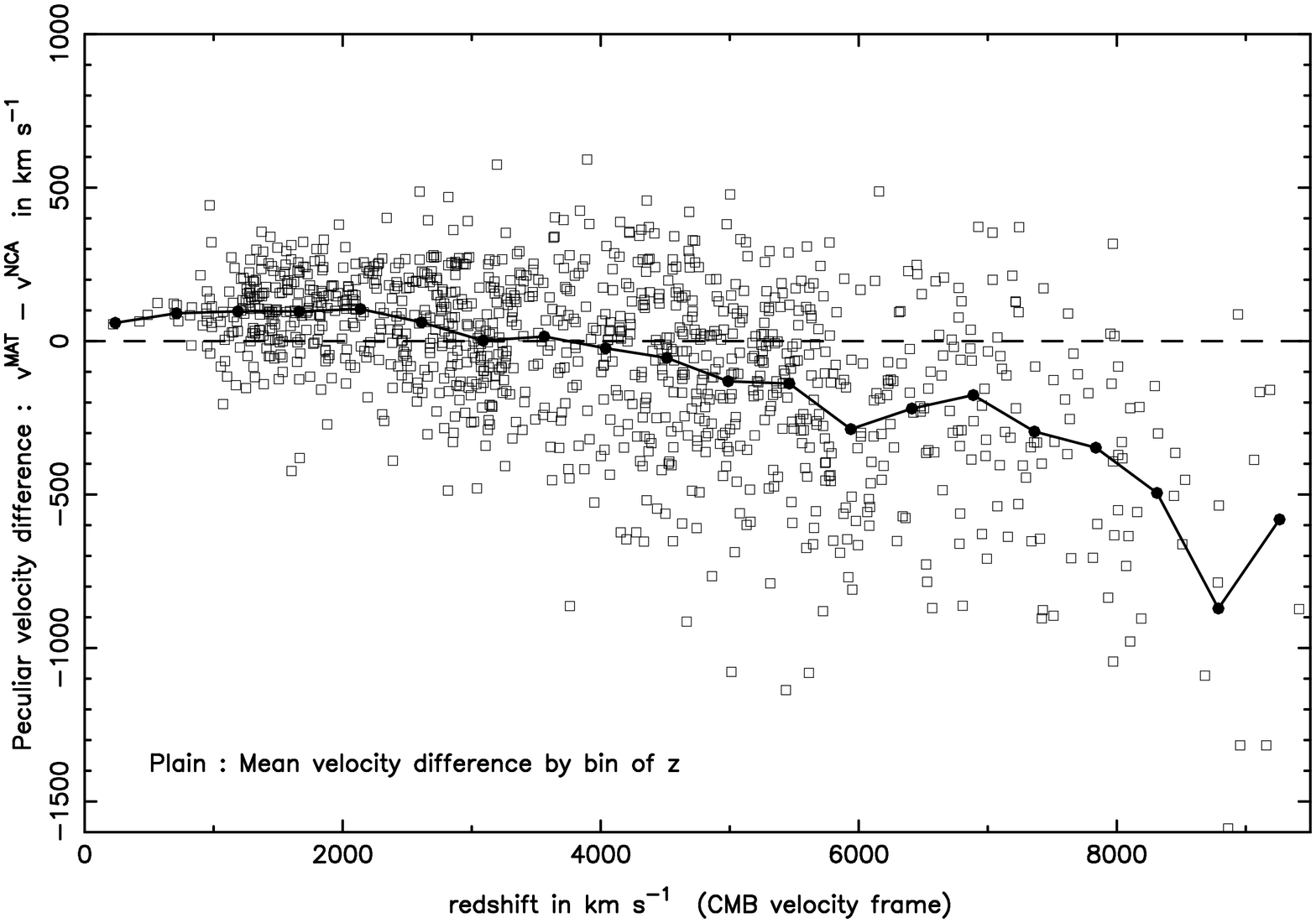,height=10.5 cm,width=\linewidth,angle=270}
\caption[]{
Top : Redshifts $z$ versus NCA distance estimates
in km s$^{-1}$ for the MAT sample (left) Outside the GA region
(right) Inside the GA region. Bottom : Peculiar velocity
difference between Mathewson velocity estimates
${\tilde v}^{\rm MAT}$ and NCA velocity estimates
${\tilde v}^{\rm NCA}$.
}
\end{figure*}

NCA calibration is herein performed by following the
discussions of section \ref{Accuracies}. The
NCA estimate of the DTF slope $a^D$ is obtained from
a subsample selected in distance
estimate beyond $4\,000$ km s$^{-1}$, which corresponds
to discard about $60\%$ of nearby galaxies
of the MAT sample. NCA estimate of the DTF "zero-point"
$B^*$ was achieved using galaxies beyond $5\,000$ km s$^{-1}$.
Figure 7 shows the results of the NCA calibration of
the Mathewson field galaxies catalog\footnote{Since
the subsampling procedure in distance estimate depends on
the values of $a^{\rm NCA}$ and $B_*^{\rm NCA}$, few iterations
were performed in order to achieve the stable situation
presented figure 7.}.

The value of NCA estimate of the DTF slope $a^D$ is
given figure 7 (bottom left) :
$a^{\rm NCA}=-7.$ with an accuracy at $r^*=4\,000$ km s$^{-1}$
given by the numerical simulations of $\sigma_a=0.15$
(or in other words $a^{\rm NCA}=-7.\pm0.15$).
Since this estimate is obtained with a large cut-off
in distance estimate, it is in principle free
of biases created by large scale peculiar velocity
field. Adding to this point that NCA calibration technique
is insentive to observational selection effects on
apparent magnitude $m$ and log line-width distance indicator $p$,
NCA estimate of DTF slope $a^D$ appears as fairly secure.

NCA estimate of the relative "zero-point" $B^*$ is shown
figure 7 (bottom right). 
The value of $B_*^{\rm NCA}=0.011$ has been arrested accounting
for the more or less stable trend  of $B_*^{\rm NCA}$ beyond
$r^*=5\,000$ km s$^{-1}$. To give a value of the
$B_*^{\rm NCA}$ accuracy is a little bit more tricky.
It was previously mentioned that the subsampling procedure
in distance estimate does not remove bias on $B_*^{\rm NCA}$
estimator created by the presence of large scale peculiar
velocity field (excepting the case of bulk flows). It turns out
that the value of this bias depends on
the specific geometry and amplitude of the real cosmic
velocity field, and so cannot be estimated without modeling this
velocity field. The amplitude of this bias for the
"Great Attractor" flow model is less 
than $\Delta B=0.001$ (or about $10\%$ of $B^*$)
for the MAT sample. It means that if the GA flow is real,
error on $B_*^{\rm NCA}$ estimator is dominated
by velocity bias rather than statistical fluctuations (accounting
for the
value of the accuracy on $B_*^{\rm NCA}$ obtained from numerical
simulations). At this stage, other criterion for
calibrating DTF relative "zero-point" $B^*$ may be
preferred, for example in constraining the average $<{\tilde v}>$
of radial peculiar velocity estimates over the sample
to vanish. Unfortunatly such property $<{\tilde v}>=0$ is
theoretically expected for fair samples (i.e. samples large enough
to be representative at any scales of the kinematical fluctutions
of the Universe), but not expected for calalogs
such as the MAT sample. 

Amplitude of velocity bias on the $B_*^{\rm NCA}$ estimator
may however be attenuated by removing of the sample
areas presumed to have a strong kinematical activity.
Such a procedure was achieved by discarding galaxies
of the MAT sample belonging to a cone pointing toward $l=310$
and $b=20$ in galactic coordinates with an angular aperture of
$45^{\rm o}$ (the GA region). This subsample (i.e. MAT sample, GA
region excluded) contains $N_{\rm gal}=799$ galaxies.
Figure 8 (left) shows
amplitudes of biases created by bulk and GA flow
when the GA region is excluded of the MAT sample.
Compared to the biases shown figure 5 (bottom left)
for the whole MAT sample, the discarding
procedure looks efficient (if of course, the "Great
Attractor" model of Bertschinger et al. \cite{Ber88}
succeeds in mimicking the real cosmic peculiar velocity field).
Figure 8 (right) shows the NCA estimate $B_*^{\rm NCA}$
for the MAT sample, GA region excluded.
If the situation improves for $r^*<3\,000$ km s$^{-1}$
(compared
to figure 6 (bottom left) showing $B_*^{\rm NCA}$
for the whole MAT sample), a residual bias persists
between $r^*=3\,000$ km s$^{-1}$ and $r^*=5\,000$ km s$^{-1}$.
The value of the NCA relative "zero-point" $B_*^{\rm NCA}=0.011$
has then to be read cautiously, keeping in mind that
it can be affected by a velocity bias (anyway
presumed not to be greater than $\Delta B=0.001$) \footnote{
Independently to these velocity
bias problems, relative "zero-point"
estimator always suffer from an additional bias inherited
from the statistical uncertainties on the slope estimate
(i.e. ${\delta B}/{B}\,\,\approx\,\, \alpha\,<p>\,\delta a^D)$}.

Finally, distance estimate ${\tilde r}$ and velocity estimate
${\tilde v}$ given respectively Eq. (\ref{distanceestimator}) and
Eq. (\ref{Betoile}) can be inferred using
the NCA calibration parameters $a^{\rm NCA}=-7.$
and $B_*^{\rm NCA}=0.011$ previously derived.
Figure 9 (left) and (right) show respectively the
redshift $z$ versus the NCA predicted distance
$H_0\,{\tilde r^{\rm NCA}} = B_*^{\rm NCA}\,\exp [\alpha
(m-a^{\rm NCA}p)]$ for galaxies of the MAT sample
outside and inside the GA region.
Figure 9 (bottom) illustrates the difference
between the peculiar velocity estimates
${\tilde v}=z- B_*\,\exp [\alpha
(m-a^D p)]$ of Mathewson et al. \cite{Mat92} and
the ones derived in this present appendix.
The averaged velocity difference by bins
of redshift $z$ is plotted function of the redshift.
It turns out that an erroneous input value of the
calibration parameters can interpret, especially at large
redshifts,  as fictitious large scale and coherent
peculiar velocity flows. The preliminar calibration step of
Tully-Fisher like relation is thus of crucial
importance for kinematical studies.

\end{document}